\documentclass[aps,floatfix,
superscriptaddress,preprint]{revtex4}

\usepackage{titlesec}
\usepackage{amssymb,amsmath,amsfonts,latexsym,graphicx,epsfig}
\usepackage{epstopdf}

\usepackage{gensymb}

\titleformat{\section}{\large\bfseries}{\thesection}{1em}{}

\newcommand{\bea}{\begin{eqnarray}}
\newcommand{\ena}{\end{eqnarray}}
\newcommand{\nn}{\nonumber\\}
\newcommand{\be}{\begin{equation}}
\newcommand{\en}{\end{equation}}

\newcommand{\ed}{\end{document}}

\newcommand{\sll}{/\kern-4pt l}
\newcommand{\slp}{p\kern-5pt/}
\newcommand{\sls}{s\kern-5pt/}
\newcommand{\slell}{/\kern-5pt\ell}
\begin{document}

\title{Probing new physics in $\bar{B}^0 \to D^{(\ast)}\tau^-\bar\nu_{\tau}$
  using the longitudinal, transverse, and normal
polarization components of the tau lepton}

\author {M. A. Ivanov}
\email{ivanovm@theor.jinr.ru}
\affiliation{
Bogoliubov Laboratory of Theoretical Physics,
Joint Institute for Nuclear Research,
141980 Dubna, Russia}

\author{J. G. K\"{o}rner}
\email{jukoerne@uni-mainz.de}
\affiliation{PRISMA Cluster of Excellence, Institut f\"{u}r Physik, 
Johannes Gutenberg-Universit\"{a}t, 
D-55099 Mainz, Germany}

\author{C. T. Tran}
\email{ctt@theor.jinr.ru,tranchienthang1347@gmail.com}
\affiliation{
Bogoliubov Laboratory of Theoretical Physics,
Joint Institute for Nuclear Research,
141980 Dubna, Russia}
\affiliation{Department of General and Applied Physics, 
Moscow Institute of Physics and Technology, 141700 Dolgoprudny, Russia}
\date{\today}

\begin{abstract}
  We study the longitudinal, transverse, and normal polarization components of the tau
  lepton in the decays $\bar{B}^0 \to D^{(\ast)}\tau^-\bar\nu_{\tau}$ and discuss
  their role in searching for new physics (NP) beyond the standard model (SM).
  Starting with a model-independent effective Hamiltonian including non-SM
  four-Fermi operators, we obtain experimental constraints on different NP
  scenarios and investigate their effects on the polarization observables. In the SM the longitudinal and transverse polarizations of the tau lepton differ substantially from the corresponding zero lepton mass values of $P_L=-1$ and $P_T=0$. In addition, $P_L$ and $P_T$ are very sensitive to NP effects. For the transverse polarization this holds true, in particular, for the effective tensor operator in the case of
  $\bar{B}^0 \to D^\ast$ and for the scalar
  operator in the case of  $\bar{B}^0 \to D$. The $T$-odd normal
  polarization $P_N$, which is predicted to be negligibly small in the SM,
  can be
  very sizable assuming NP complex Wilson coefficients. We also discuss in some
  detail how the three polarization components of the tau lepton can be measured
  with the help of its  subsequent leptonic and semihadronic decays. 
\end{abstract}


\maketitle

\section{Introduction}
\label{sec:intro} 
The exclusive semileptonic decays $\bar{B}^0 \to D^{(\ast)}\tau^-\bar\nu_{\tau}$ have
been measured by the {\it BABAR}~\cite{Lees:2012xj}, Belle~\cite{Huschle:2015rga,Sato:2016svk}, and LHCb~\cite{Aaij:2015yra} collaborations in an effort to
unravel the well-known $R_{D^{(*)}}$ puzzle  which has persisted for several
years~\cite{Ivanov:2016qtw,Crivellin:2012ye,Becirevic:2012jf,DeVito:2004zs,Nierste:2008qe,Faustov:2012nk,Celis:2012dk,Biancofiore:2013ki,Duraisamy:2013kcw,Sakaki:2013bfa,Duraisamy:2014sna,Calibbi:2015kma,Greljo:2015mma,Bauer:2015knc,Fajfer:2015ycq,Barbieri:2015yvd,Barbieri:2016las,Boucenna:2016wpr,Li:2016vvp,Nandi:2016wlp,Alok:2016qyh,Deshpand:2016cpw,Kim:2016yth,Bardhan:2016uhr,Dutta:2016eml,Wang:2016ggf,Celis:2016azn,Bhattacharya:2016zcw,Becirevic:2016yqi,Becirevic:2016oho}. Recently, the Belle collaboration reported a new measurement of the decay
$\bar{B}^0 \to D^\ast\tau^-\bar\nu_{\tau}$ using the hadronic $\tau^-$ decay modes $\tau^{-}\to \pi^-\nu_{\tau}$ and $\tau^{-}\to\rho^-\nu_{\tau}$, in which
they found $R_{D^\ast} = 0.270 \pm 0.035(\text{stat.})^{+0.028}_{-0.025}(\text{syst.})$~\cite{Hirose:2016wfn}. Taking this new result into account, the current
world averages of the ratios are $R_D = 0.406 \pm 0.050$ and
$R_{D^\ast} = 0.311 \pm 0.016$, which exceed the SM predictions of
$R_D = 0.300 \pm 0.008$~\cite{Na:2015kha,Lattice:2015rga,Aoki:2016frl} and $R_{D^\ast}
= 0.252 \pm 0.003$~\cite{Fajfer:2012vx} by $2.1\sigma$ and $3.6\sigma$, respectively. 

In Ref.~\cite{Hirose:2016wfn} the Belle collaboration also reported
on the first measurement of the longitudinal polarization of the tau lepton
in the decay
$\bar{B}^0 \to D^\ast\tau^-\bar\nu_{\tau}$ with the result $P_L^\tau = -0.38 \pm 0.51(\text{stat.})
^{+0.21}_{-0.16} (\text{syst.})$. The errors of this measurement are quite
large
but this pioneering measurement has opened a
completely new window on the analysis of the dynamics of the
semileptonic $B\to D$ and $B \to D^*$ transitions. The hope is that, with
the Belle II super-B factory nearing completion,
more precise values of the polarization can be achieved in the future, which
would shed more light on the
search for possible NP in these decays.

In this paper we shall study the longitudinal, transverse, and normal
polarization components of the $\tau^-$ in the semileptonic decays
$\bar{B}^0 \to D^{(\ast)}\tau^-\bar\nu_{\tau}$. In order to set up our notation we define three orthogonal unit vectors as follows:
\be
\vec{e}_L=\frac{\vec{p}_\tau}{|\vec{p}_\tau|},\qquad \vec{e}_N=\frac{\vec{p}_\tau \times\vec{p}_{D^{(*)}}}{|\vec{p}_\tau \times\vec{p}_{D^{(*)}}|},
\qquad \vec{e}_T=\vec{e}_N\times\vec{e}_L,
\en
where $\vec{p}_\tau$ and $\vec{p}_{D^{(*)}}$ are the three-momenta of the $\tau^-$ and the $D^{(*)}$ meson in the $W^-_{off-shell}$ rest frame. In the following we
shall loosely refer to this frame as the $W^-$ rest frame. The three unit vectors
$\vec{e}_T$, $\vec{e}_N$, and $\vec{e}_L$ form a right-handed coordinate system. The longitudinal $(L)$, normal $(N)$, and transverse $(T)$ polarization four-vectors of the $\tau^-$ in its rest frame are given by
\be
s^\mu_L=(0,\vec{e}_L),
\qquad s^\mu_N=(0,\vec{e}_N),
\qquad s^\mu_T=(0,\vec{e}_T).
\en
A Lorentz boost from
the $\tau^-$ rest frame to the $W^-$ rest frame transforms only the longitudinal polarization four-vector according to
\be
s^\mu_L=\Big(\frac{|\vec{p}_\tau|}{m_\tau},\frac{E_\tau}{m_\tau}\frac{\vec{p}_\tau}{|\vec{p}_\tau|}\Big),
\en
leaving the normal $(s^\mu_N)$ and transverse $(s^\mu_T)$ polarization four-vectors unchanged.
The longitudinal, normal, and transverse polarization components of the tau
are given by
\be
\label{eq:poldef}
P_i(q^2) = \frac{d\Gamma(s^\mu_i)/dq^2-d\Gamma(-s^\mu_i)/dq^2}{d\Gamma(s^\mu_i)/dq^2+d\Gamma(-s^\mu_i)/dq^2},\qquad i=L, N, T,
\en
where $q^\mu=p_B^\mu-p_{D^{(*)}}^\mu$ is the momentum transfer. We note that the terms longitudinal polarization and longitudinal polarization component are often used interchangeably, as in this paper. The same convention is used for the normal and transverse polarizations.

The normal polarization component $P_N$ is a $T$-odd observable and is predicted
to be zero in the SM in the absence of final state interactions which are known
to be negligibly small. However, in some extended versions of the SM such
as the two-Higgs-doublet models, the minimal supersymmetric standard model, and the leptoquark model, large values of $P_N$ are
possible through the introduction of $CP$-violating
phases~\cite{Garisto:1994vz, Tsai:1996ps, Wu:1997uaa, Lee:2001nw}.

The longitudinal polarization  $P_L$ has also been used as a promising observable in
order to probe NP in $\bar{B}^0 \to D^{(\ast)}\tau^-\bar\nu_{\tau}$~\cite{Tanaka:2010se, Fajfer:2012vx, Datta:2012qk,Tanaka:2012nw, Bhattacharya:2015ida, Becirevic:2016hea}. $P_L$ has been found to be very sensitive to the scalar and tensor operators. It has been shown in Ref.~\cite{Tanaka:2010se, Tanaka:2012nw} that some correlations between $P_L$ and the decay rate are very useful for NP prediction. In addition, the NP couplings can be extracted from $P_L$ with much less uncertainties as compared to those from other observables~\cite{Bhattacharya:2015ida}.

In Ref.~\cite{Ivanov:2015tru} we have calculated the SM values of the longitudinal
and transverse polarization of the charged lepton in the decays
$\bar{B}^0 \to D^{(\ast)}\ell^-\bar\nu_{\ell}$. The polarization components have been
calculated in the so-called helicity basis where the  polarization components
are given in terms of bilinear forms of the helicity amplitudes of the
current-induced
$B \to D^{(\ast)}$ transitions. Depending on the phase space region the
transverse
tau polarization can become quite large.
On average one has $<P^\tau_T>=0.84$ ($B\to D$) and $<P^\tau_T>=0.46$  
($B\to D^{\ast}$)~\cite{Ivanov:2015tru} compared to $<P^\ell_T>=0$  for
$m_{\ell}=0$ in both cases.
For the longitudinal polarization one has $<P^\tau_L>=0.33$ ($B\to D$) and 
$<P^\tau_L>=-0.50$ ($B \to D^{\ast}$)~\cite{Ivanov:2015tru}
which one has to compare with the zero lepton mass result
$<P^\ell_L>=-1$, again in both cases \footnote{One of the authors of the present
  paper (J.G.K.) records his contentment that the values
  $<P^\tau_L>=0.33$ ($B\to D$) and $<P^\tau_L>=-0.53$ ($B \to D^{\ast})$ calculated
  in the early paper~\cite{Korner:1989qb} are quite close to those calculated in
~\cite{Ivanov:2015tru}.}. For the averages of the total polarization
$|\vec P^\tau|$ one obtains $<|\vec P^\tau|>=0.91$ ($B\to D$) and
$<|\vec P^\tau|>=0.71$ ($B \to D^{\ast})$. In this paper we also consider the transverse polarization in the presence of NP and compare its NP sensitivity with that of $<P^\tau_L>$ and $<P^\tau_N>$. The discussion of NP contributions to the
transverse and normal polarization components of the $\tau^-$ are new.

Since the $\tau^-$ lepton decays weakly, its polarization is revealed through its ensuing decay distributions, i.e. it is self-analyzing. As analyzing modes for the $\tau^{-}$ polarization we will consider the 
four dominant $\tau^{-}$ decay modes
\bea
\label{eq:taumode}
\tau^{-}&\to& \pi^{-}\nu_{\tau} \quad (10.83\%),\qquad
\tau^{-}\to \mu^{-}\bar \nu_{\mu}\nu_{\tau} \quad (17.41\%),\nn
\tau^{-}&\to& \rho^{-}\nu_{\tau}\quad (25.52\%)
,\qquad
\tau^{-}\to e^{-}\bar \nu_{e}\nu_{\tau} \quad(17.83\%),
\ena
where we have added the respective branching fractions in brackets. In the next
section, we will show how the three polarization components of the tau can be
measured by using its decays as polarization analyzers and how well each mode
can serve as polarization analyzer. The remaining parts of the paper
are organized as follows: in Sec.~\ref{sec:formalism} we introduce some formalism concerning the semileptonic transitions, including the derivation of the polarization formulae in the presence of NP. An analysis of NP effects on the polarizations is given in Sec.~\ref{sec:analysis}. Finally, we summarize the main results in Sec.~\ref{sec:summary}.
\section{Analyzing the polarization of the tau through its decays }
\label{sec:tau-decays}
\begin{figure}[htbp]
\includegraphics[scale=0.4]{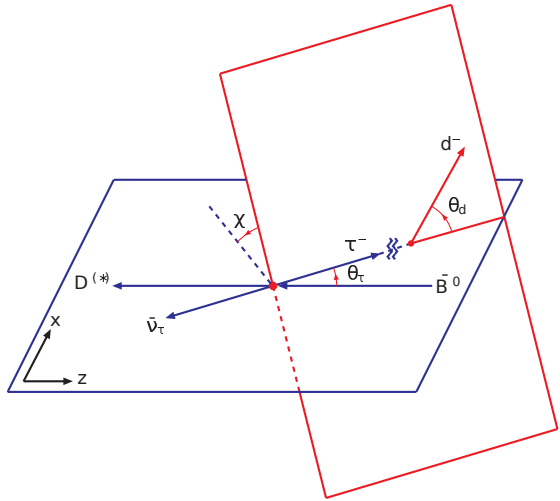}
\caption{Kinematics of the decay $\bar{B}^0 \to D^{(\ast)}\tau^-\bar\nu_{\tau}$ followed by a $\tau^-$ decay. See text for more details.}
\label{fig:angle}
\end{figure}

The polarization components of the $\tau^-$ in $\bar{B}^0 \to D^{(\ast)}\tau^-\bar\nu_{\tau}$ can be measured by using the decay products of the $\tau^-$ as polarization
analyzers. The kinematics of the decay $\bar{B}^0 \to D^{(\ast)}\tau^-\bar\nu_{\tau}$ followed by a $\tau^-$ decay is depicted in Fig.~\ref{fig:angle}, where $d^-=\pi^-,\,\rho^-,\,e^-,\,\mu^-$.
In the $W^-$ rest frame, $\theta_\tau$ is the angle between the $\tau^-$ three-momentum and the direction opposite to the direction of the $D^{(*)}$ meson. In the $\tau^-$ rest frame, $\theta_d$ is the angle between the three-momentum of the final tau daughter $d^-$ and the longitudinal polarization axis which is chosen to coincide with the direction of the $\tau^-$ in the $W^-$ rest frame (helicity basis). The production plane defined by the decay $\bar{B}^0 \to D^{(\ast)}\tau^-\bar\nu_{\tau}$ is spanned by the three-momenta of the $\tau^-$ and the $D^{(\ast)}$ while the $\tau^- \to d^-+X$ decay plane is spanned by the three-momentum of the $d^-$ and the longitudinal polarization axis. The angle $\chi$ is the azimuthal angle between the two planes. We choose a right-handed $xyz$ coordinate system in the $W^-$ rest frame such that the $z$ axis is opposite to the direction of the mesons $\bar{B}^0$ and $D^{(\ast)}$, and the three-momentum of the $\tau^-$ lies in the $(xz)$ plane. In this system the $\tau^-$ momentum is given by 
\be
p_\tau^\mu=E_\tau(1, \beta_\tau\sin\theta_\tau,0,\beta_\tau\cos\theta_\tau),
\en
where $E_\tau=(q^2+m_\tau^2)/2\sqrt{q^2} $ is the energy and $\beta_\tau=|\vec{p}_\tau|/E_\tau=\sqrt{1-m_\tau^2/E_\tau^2}$ the velocity of the $\tau^-$ in the $W^-$ rest frame.

Let us discuss the spin-momentum correlations in the $\tau^-$ rest frame. Since we are dealing with two-body
decays ($\tau^- \to \pi^-(\rho^-) +\nu_\tau$) or quasi-two-body decays
($\tau^- \to \ell^- +X$) there is only one independent spin-momentum scalar
product which can be taken to be $(\vec p_d \cdot \vec P)$, where $\vec p_d$ is the three-momentum of the $d^-$ and $\vec P$ is the polarization vector of the $\tau^-$. Note that in a three-body decay as e.g. in $t \to b+\ell^+ +\nu_\ell$ there are two possible spin-momentum scalars which provide for a richer spin-momentum correlation structure (see e.g.~\cite{Korner:1998nc,Groote:2006kq}). Returning to the two-body decays treated here the differential
polar angle distribution is given by
\be
\label{eq:angle-dp}
\frac{d\Gamma}{dq^2d\cos\theta_{dP}}={\cal B}_d\frac{d\Gamma}{dq^2}
\frac{1}{2}\,\Big(1+A_d \cdot |\vec P(q^2)| \cos\theta_{dP}\Big),
\en
where $\theta_{dP}$ is the polar angle between the momentum $\vec p_d$ and
the polarization vector $\vec P$ of the $\tau^-$, and ${\cal B}_d$ and $A_d$ are the branching fraction
and the analyzing power of the decay $\tau^- \to d^-+X$, respectively. Note that the magnitude of the analyzing power
has to satisfy $|A_d| \le 1$ to guarantee the positivity of rates for
$|\vec P|=1$.

The polar angle $\theta_{dP}$ appearing in Eq.~(\ref{eq:angle-dp}) is
experimentally not accessible since the direction of the polarization
vector $\vec P$ of the
$\tau^-$ is {\it a priori} unknown. However, one can define experimentally
accessible angles $\theta_d$ and $\chi$ through the representation of the
momentum vector $\vec p_d$ in the production plane (see Fig.~\ref{fig:angle}) via
\be
\vec p_d = |\vec p_d| (\sin\theta_d \cos\chi, \sin\theta_d \sin\chi,
\cos\theta_d).
\en
In terms of the angles $\theta_d$ and $\chi$, the decay distribution reads
\be
\label{eq:ang-distr}
\frac{d\Gamma}{dq^2d\cos\theta_{d}d\chi/2\pi}
={\cal B}_ d\frac{d\Gamma}{dq^2}
\frac{1}{2}\big[1+A_d ( P_T(q^2) \sin\theta_d \cos\chi
+P_N(q^2) \sin\theta_d \sin\chi +P_L(q^2)\cos\theta_d)  \big].
\en
Through an analysis of the decay distribution~(\ref{eq:ang-distr}) one can
determine the three components of the $q^2$-dependent polarization vector 
$\vec P(q^2)=(P_T(q^2),P_N(q^2),P_L(q^2))$.

Upon $\chi$ integration, one obtains
\be
\label{eq:theta-distr}
\frac{d\Gamma}{dq^2d\cos\theta_{d}}
={\cal B}_ d\frac{d\Gamma}{dq^2}\frac{1}{2}\Big(1+A_d  P_L(q^2)\cos\theta_d   \Big) 
\en
such that the forward-backward polarization asymmetry is given by
\be
\label{eq:FBpol}
A_{FB}^P=\frac{d\Gamma(F)-d\Gamma(B)}{d\Gamma(F)+d\Gamma(B)}
=A_d P_L(q^2).
\en
Upon $\cos\theta_{d}$ integration one has
\be
\label{eq:chi-distr}
\frac{d\Gamma}{dq^2d\chi/2\pi}
={\cal B}_ d\,\frac{d\Gamma}{dq^2}\Big(1+A_d \frac{\pi}{4} \big( P_T(q^2)
\cos\chi+P_N(q^2) \sin\chi \big)  \Big)
\en
with an effective azimuthal analyzing power of $A_d\pi/4$.
\subsection{The semihadronic modes $\tau^{-} \to \pi^{-} \nu_{\tau} $
 and $\tau^{-} \to  \rho^{-} \nu_{\tau} $ }
The differential decay rate of $\bar{B}^0 \to D^{(\ast)}\tau^-(\to \pi^{-} \nu_{\tau})\bar\nu_{\tau}$ reads
\be
\label{3foldpi}
\frac{d\Gamma_\pi}{dq^{2}d\cos\theta_\pi d\chi/2\pi}
={\cal B}_{\pi}\frac{d\Gamma}{dq^2}\frac{1}{2} \big[1+P_T(q^2) \sin\theta_\pi \cos\chi
+P_N(q^2) \sin\theta_\pi \sin\chi+ P_L(q^2) \cos\theta_\pi\big],
\en
where ${\cal B}_{\pi}$ is the branching fraction of $\tau^{-}\to \pi^{-} \nu_{\tau}$ and $\Gamma$ is the decay rate of $\bar{B}^0 \to D^{(\ast)}\tau^-\bar\nu_{\tau}$. Note
that the analyzing power of the decay $\tau^{-}\to \pi^{-} \nu_{\tau}$ is
$100\%$. In the following
we shall drop explicit reference to the  component $P_N$ in the angular decay
distribution. After $\cos\theta_\pi$ integration, one obtains 
\be
\label{2foldpi}
\frac{d\Gamma_\pi}{dq^{2}d\chi/2\pi}
={\cal B}_{\pi} \frac{d\Gamma}{dq^2}\Big(1+\frac{\pi}{4}P_T(q^2)\cos\chi\Big).
\en
The effective azimuthal analyzing power is quite large with $\pi/4=78.54\%$. 

For the decay $\bar{B}^0 \to D^{(\ast)}\tau^-(\to \rho^{-} \nu_{\tau})\bar\nu_{\tau}$ one has
\be
\label{rho}
\frac{d\Gamma_{\rho}}{dq^{2}d\cos\theta_\rho d\chi/2\pi}=
{\cal B}_{\rho}\frac{d\Gamma}{dq^2}\frac{1}{2}\Big[1+\frac{m_{\tau}^{2}-2m_{\rho}^{2}}{m_{\tau}^{2}+2m_{\rho}^{2}}(P_T(q^2) \sin\theta_\rho \cos\chi
+ P_L(q^2) \cos\theta_\rho)
\Big],
\en
where ${\cal B}_{\rho}$ is the branching fraction of
$\tau^{-} \to \rho^{-} \nu_{\tau} $. It is apparent that one looses analyzing
power compared to the case
$\tau^{-}\to \pi^{-} \nu_{\tau} $
since $(m_{\tau}^{2}-2m_{\rho}^{2})/(m_{\tau}^{2}+2m_{\rho}^{2})=0.4485<1$.

However, one can retain the full analyzing power if one
projects out the longitudinal and transverse components of the
$\rho^-$, which can be achieved by an angular analysis of the decay
$\rho^- \to \pi^-+\pi^0$ in the rest frame of the $\rho^-$. The polar angle
distribution of the decay $\rho^- \to \pi^-+\pi^0$ reads
\be
\frac{d\Gamma_{\rho}}
{d\cos\theta}=\tfrac 38(1+\cos^2\theta)\Gamma^T+\tfrac 34 \sin^2\theta\Gamma^L,
\en
where $\theta$ is the polar angle of the $\pi^-$ with respect to the original flight
direction of the $\rho^-$. Technically,
one can project out the longitudinal piece of the $\rho^-$ with the help of
the normalized
longitudinal polarization four-vector of the $\rho^-$ which reads
\be
\varepsilon^{\alpha}(0)
= \frac{1}{m_{\rho}m_{\tau}p}
(m_{\rho}^{2}p_{\tau}^{\alpha}-p_{\rho}p_{\tau}p_{\rho}^{\alpha}).
\en
One can check that $p_{\rho}\cdot\varepsilon(0)=0$ and that the 
polarization four-vector is correctly normalized: 
$\varepsilon^{\ast}(0)\cdot\varepsilon(0)=-1$. In the rest frame of the $\rho^{-}$
one has $\varepsilon^{\alpha}(0)=(0;0,0,1)$.
The transverse 
contribution can be obtained from the difference 
$\Gamma^{T}=\Gamma-\Gamma^{L}$.

The longitudinal and transverse differential decay distributions of the $\rho^-$
are finally given by
\bea
\label{TandL}
\frac{d\Gamma_{\rho}^{L}}
{dq^{2}d\cos\theta_\rho d\chi/2\pi}&=&
{\cal B}_{\rho}\frac{d\Gamma}{dq^2}\frac{m_{\tau}^{2}/2}{m_{\tau}^{2}+2m_{\rho}^{2}}
\Big[1+\big(P_T(q^2) \sin\theta_\rho \cos\chi
+ P_L(q^2) \cos\theta_\rho\big)
\Big],\nn
\frac{d\Gamma_{\rho}^{T}}
{dq^{2}d\cos\theta_\rho d\chi/2\pi}&=&
{\cal B}_{\rho}\frac{d\Gamma}{dq^2}\frac{m_{\rho}^{2}}{m_{\tau}^{2}+2m_{\rho}^{2}}
\Big[1-\big(P_T(q^2) \sin\theta_\rho \cos\chi
+ P_L(q^2) \cos\theta_\rho\big)
\Big].
\ena
By separating the two distributions on has regained the full
analyzing power of $100\%$ in both cases. This can e.g. be done by projection:
${\cal P}^L=2(1- 5/2\cos^2\theta)$ will project out the longitudinal and
${\cal P}^T=-(1- 5\cos^2\theta)$ the transverse component. It is evident that the
sum of the two distributions~(\ref{TandL}) gives the result Eq.~(\ref{rho}).

\subsection{The leptonic modes 
  $\tau^{-} \to \ell^{-}\bar \nu_{\ell}\nu_{\tau}$ $(\ell=e,\mu)$ }

Using the results of e.g. Ref.~\cite{Fischer:2002hn} one finds
\be
\label{leptonic}
\frac{d\Gamma_\ell}{dq^2dxd\cos\theta_\ell d\chi/2\pi}=
\frac{d\Gamma}{dq^2}\frac{\Gamma_0}{\Gamma_\tau}\beta x \big[G_{1}(x)
  +G_{2}(x)\big(P_T(q^2) \sin\theta_\ell \cos\chi
+ P_L(q^2) \cos\theta_\ell\big)
\big],
 \en
 where, as usual, we have defined a scaled energy variable
 $x=2E/m_\tau$ where $E=(|\vec{p_\ell}|^{2}+m_{\ell}^{2})^{1/2}$ is the energy of the final charged lepton $\ell^-$
 in the $\tau^{-}$ rest frame. Here, 
$\Gamma_0=G_F^2m_\tau^5/192\pi^3$ is the reference rate for the leptonic
 decay of final-state massless leptons $m_\ell=0$, and $\Gamma_\tau$ is the total decay width of the $\tau^-$.  Note that the
 expression to the right of $\Gamma_0/\Gamma_\tau$ integrates to 1 for $m_\ell=0$
 as it should be.
 For later purposes we define a reference branching ratio
 ${\cal B}_\ell^0=\Gamma_0/\Gamma_\tau $.
 
The coefficient functions in Eq.~(\ref{leptonic}) are given by~\cite{Fischer:2002hn}
\be
G_{1}=x(3-2x)-(4-3x)y^{2}, \qquad G_{2}=\beta x(1-2x+3y^{2}),
\en
where $y=m_\ell/m_\tau$ and
$\beta = \sqrt{1-4y^2/x^2}
= \sqrt{1-m_{\ell}^2/E^2}= p/E$. We mention that the next-to-leading order QED radiative corrections
to the leptonic polarized $\tau^-$ decays can also be found
in Ref.~\cite{Fischer:2002hn}.

The polar and azimuthal analyzing power is determined by the ratio $G_2(x)/G_1(x)$.
By averaging over $x$ $(2y\leq x\leq 1+y^2)$, one obtains
\be
\frac{<\beta x G_2(x)>}{<\beta x G_1(x)>}=-\tfrac {1}{12}(1+8y^2-32y^3 +\dots).
\en
The azimuthal analyzing power is given by
\be
\label{aziana}
\frac{d\Gamma_\ell}{dq^2d\chi/2\pi}=
\frac{d\Gamma}{dq^2}{\cal B}_\ell^0 (1+P_T A_\chi \cos\chi), \qquad {\rm where} \quad
A_\chi= -\frac{\pi}{12}(1+8y^2-32y^3 +\dots).
\en
For $m_\ell=0$ one finds $A_\chi=-0.262$ which increases by $3.2\%$ for
$m_\ell=m_\mu$ [see Eq.~(\ref{aziana})].

Another possibility to analyze the polarization of the $\tau^-$ is to describe the
leptonic decay of the polarized $\tau^-$ in terms of the variables $(x,\,z)$,
where $z=E_{\ell}/E_{\tau}$ is the fractional energy $E_{\ell}$ of the daughter lepton
 and the energy $E_{\tau}$ of the $\tau^-$ 
both in the $W^-$ rest frame~\cite{Bullock:1992yt}. For the dependence
$z=z(x,\,\cos\theta_\ell)$,
one finds
\be
z=\frac{E_{\ell}}{E_{\tau}}= \frac{\beta_{\tau}p\cos\theta_\ell +E}{m_{\tau}}
=\frac{x}{2}(\beta_{\tau}\beta \cos\theta_\ell +1).
\en
It is important to realize that $E$ (energy of the daughter lepton in the $\tau^{-}$ rest
frame) is no longer fixed but becomes a variable to be integrated over.

Let us first discuss the so-called collinear approximation
$\beta_{\tau}=1$ and the zero lepton mass limit $\beta=1$ introduced in Ref.~\cite{Bullock:1992yt}
to analyze the longitudinal polarization of the
$\tau^-$.
The approximation $\beta_{\tau}=1$ is good for the small recoil (i.e. large $q^2$) 
region. The approximation $\beta=1$ holds for the limiting case when one
can neglect the lepton mass in the final state.
With these approximations the twice differential rate reads
\be
\frac{d\Gamma_\ell}{dq^2dxdz d\chi/2\pi}
= {\cal B}_{\ell}^0\frac{d\Gamma}{dq^2}2 \big(G_{1}(x)+G_{2}(x)
(P_{L}\cos\theta_\ell+P_{T}\sin\theta_\ell \cos\chi)\big).
\label{eq:2approx}
\en
By integrating Eq.~(\ref{eq:2approx}) over $x$
in the region $z \leq x \leq 1$, one obtains
\bea
\label{eq:coll}
\frac{d\Gamma_\ell}{dq^2 dz d\chi/2\pi}
&=& {\cal B}_{\ell}^0\frac{d\Gamma}{dq^2}\tfrac{1}{3}(1-z) \big[(5+5z-4z^{2})+P_L(q^2)(1+z-8z^{2})\nn
&&-\tfrac 85P_T(q^2)\sqrt{z(1-z)}(1+4z)\cos\chi\big].
\ena
The differential rate and the contribution proportional to $P_L$ agree 
with Eq.~(2.8) of Ref.~\cite{Bullock:1992yt}. Upon $z$ integration
$(0 \leq z \leq 1)$, one obtains the azimuthal distribution
\be
\frac{d\Gamma_\ell}{dq^2 d\chi/2\pi}
={\cal B}_{\ell}^0\frac{d\Gamma}{dq^2} \Big(1-\frac{\pi}{12} 
P_T(q^2)\cos\chi\Big).
\en
The analyzing power is $\pi/12=26.18\%$, which is in agreement with the corresponding
result in Eq.~(\ref{aziana}).

The calculation for $\beta_{\tau} \neq 1$ and $\beta=1$ is slightly more
difficult and has been done
by Tanaka and Watanabe~\cite{Tanaka:2010se} for the differential rate and
the longitudinal contribution 
proportional to $P_{L}$. 
The decay
distribution in terms of $dz$ is written as
\be
\frac{d\Gamma_\ell}{dq^2 dz d\chi/2\pi}
= {\cal B}_{\ell}\frac{d\Gamma}{dq^2} \big[f(q^2,z)+g(q^2,z)P_L(q^2)+h(q^2,z)P_T(q^2)\cos\chi],
\en
where ${\cal B}_{\ell}$ is the branching fraction of $\tau^{-}\to \ell^{-}\bar \nu_{\ell}\nu_{\tau}$.
Neglecting the lepton mass $m_\ell$, i.e. setting $\beta=1$, the functions $f$, $g$, and $h$ are given by
\bea
\label{eq:noncoll1}
f(q^2,z) &=& \frac{16z^2}{3(1-\beta_\tau^2)^3}\big[9(1-\beta_\tau^2)-4(3+\beta_\tau^2)z\big],\nn
g(q^2,z) &=& -\frac{16z^2}{3(1-\beta_\tau^2)^3}\beta_\tau\big[3(1-\beta_\tau^2)-16z\big],\\
h(q^2,z) &=& \frac{4\pi z^2}{(1-\beta_\tau^2)^3}\sqrt{1-\beta_\tau^2}(1-\beta_\tau^2-4z),\nonumber
\ena
for $0\leq z\leq (1-\beta_\tau)/2$, and
\bea
\label{eq:noncoll2}
f(q^2,z) &=& \frac{1+\beta_\tau -2z}{3\beta_\tau(1+\beta_\tau)^3}\big[5(1+\beta_\tau)^2+10(1+\beta_\tau)z-16z^2\big],\nn
g(q^2,z) &=& \frac{1+\beta_\tau -2z}{3\beta_\tau^2(1+\beta_\tau)^3}\big[(1+\beta_\tau)^2+2(1+\beta_\tau)z-8(1+3\beta_\tau)z^2\big],\\
h(q^2,z) &=& \frac{4z^2}{(1-\beta_\tau^2)^3}\sqrt{1-\beta_\tau^2}(1-\beta_\tau^2-4z)\Big(\frac{\pi}{2}+\arcsin\frac{1-\beta_\tau^2-2z}{2z\beta_\tau}\Big)\nn
&&-\frac{\sqrt{\beta_\tau^2-1+4z-4z^2}}{3\beta_\tau^2(1-\beta_\tau^2)^2}\big[(1-\beta_\tau^2)^2+2(1-\beta_\tau^2)z-8(1+2\beta_\tau^2)z^2\big],\nonumber
\ena
for $(1-\beta_\tau)/2 \leq z \leq (1+\beta_\tau)/2$. Equations.~(\ref{eq:noncoll1}) and~(\ref{eq:noncoll2}) are obtained by integrating Eq.~(\ref{eq:2approx}) over $x$ in the regions $2z/(1+\beta_\tau) \leq x \leq 2z/(1-\beta_\tau)$ and $2z/(1+\beta_\tau) \leq x \leq 1$, respectively.

In the collinear approximation 
$\beta_\tau=1$, the first region
$0\leq z\leq (1-\beta_\tau)/2$ shrinks to zero, while the second region  $(1-\beta_\tau)/2\leq z\leq (1+\beta_\tau)/2$ simplifies to $0\leq z\leq 1$. The collinear forms of the functions $f(q^2,z)$ and $g(q^2,z)$ in Eq.~(\ref{eq:coll}) can be obtained by simply substituting $\beta_\tau=1$ in Eq.~(\ref{eq:noncoll2}). However, it is quite subtle to recover the
collinear form of $h(q^2,z)$ in Eq.~(\ref{eq:coll}) from Eq.~(\ref{eq:noncoll2}) since the treatments of the integral in two cases are different, depending on whether $\beta_\tau=1$ or $\beta_\tau\neq1$.

Yet another method to analyze the longitudinal polarization of the $\tau^-$
has been suggested in Ref.~\cite{Alonso:2016gym} where a forward-backward asymmetry
is defined with respect to $\cos\theta^\ast$, where $\theta^\ast$ is the angle
between the final charged lepton and the recoiling $D^{(*)}$ in the
$W^-$ rest system. At the end of this section
we shall also discuss a different basis, the so-called off-diagonal basis,
where the $z$ axis is chosen to point in the direction of the polarization
vector of the $\tau^-$.

\subsection {The off-diagonal basis}
In their papers \cite{Mahlon:1995zn,Parke:1996pr,Mahlon:1997uc}
 Mahlon, Parke, and Shadmi introduced the so-called off-diagonal (OD)
basis to maximize spin-spin correlation effects in top quark pair production
in $e^+e^-$ and hadronic interactions. As shown in Ref.~\cite{Groote:2010zf,Kaldamae:2016gws}
the off-diagonal basis amounts to choosing the $z$ axis to point in the
direction
of the polarization vector of the top quark, or, in this application, of the
polarization vector of the $\tau^-$. For the sake of simplicity, we shall only discuss
the off-diagonal basis for the SM case where $P_N=0$.

The relevant rotation to the off-diagonal basis is achieved by a rotation
in the $(\vec e_L,\vec e_T)$ plane by an angle $\theta_{OD}$ where $\theta_{OD}$ is the polar angle of the tau polarization relative to the tau three-momentum, measured anticlockwise from the direction of the tau. One has
\be
\frac{\sin\theta_{OD}}{\cos\theta_{OD}}=\frac{P_T}{P_L}.
\en
In the off-diagonal basis (denoted by a prime), the transverse component of the
polarization vector is zero $P^\prime_T=0$ and the azimuthal
contribution proportional
to $\cos \chi$ in the angular decay distributions vanishes. Therefore,
the sensitivity of the polar angle measurement proportional to $\cos\theta^\prime_d$ is enhanced since $|\vec P|=\sqrt{P_T^2+P_L^2} \ge |P_L|$. Here, $\theta^\prime_d$ is the polar angle between the three-momentum of the $d^-$ and the $z$ direction in the off-diagonal basis.

This discussion suggests a possible search strategy to experimentally determine
the polarization vector of the $\tau^-$ from a set of polar measurements
alone. Take a set of
directions $z$ in the
$(\vec e_L,\vec e_T)$ plane and maximize the forward-backward polarization
asymmetry $A_{FB}^P=A_d P_L(q^2)$ for this set. The $z$ direction corresponding
to this maximal value gives the direction of the $\tau^-$ polarization vector $\vec P$, and the corresponding
value of $P_L(q^2)$ obtained from $A_{FB}^P=A_d P_L(q^2)$ determines its
magnitude $|\vec P|$.

\begin{figure}[htbp]
\includegraphics[scale=0.6]{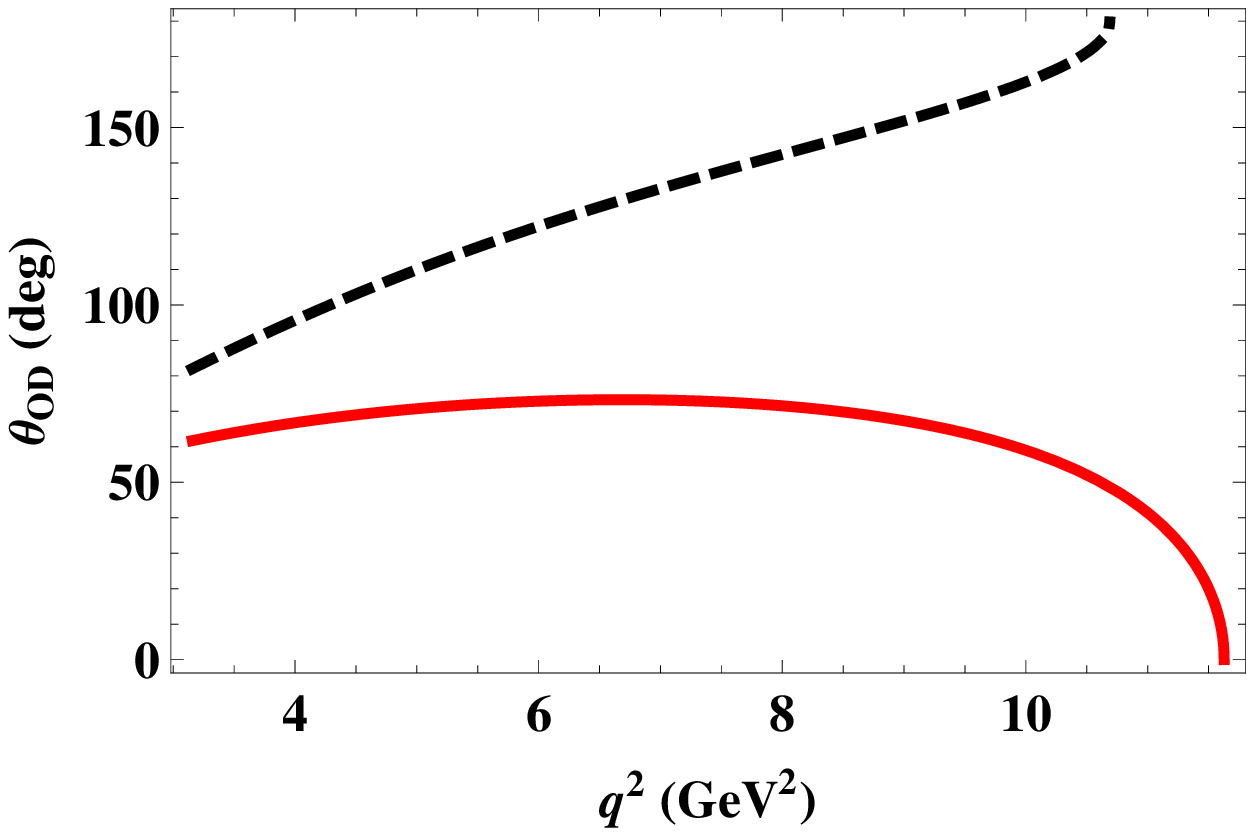}
\caption{The angle $\theta_{OD}$ for the $B\to D^\ast$  (dashed line) and $B\to D$ (solid line) transitions.}
\label{fig:OD_angle}
\end{figure}

In Fig.~\ref{fig:OD_angle} we display the $q^2$ dependence of the angle $\theta_{OD}$ for the $B\to D$ and $B\to D^\ast$ transitions. In the case of the  $B\to D$ transition the angle $\theta_{OD}$ slightly changes in the range $(50\degree,70\degree)$ for almost the whole $q^2$ region and quickly decreases from $50\degree$ to $0\degree$ for $q^2 \gtrsim 10~\text{GeV}^2$. In the case of the  $B\to D^\ast$ transition the angle $\theta_{OD}$ monotonically increases with  $q^2$ from about $80\degree$ to $180\degree$.

\begin{figure}[htbp]
\includegraphics[scale=0.65]{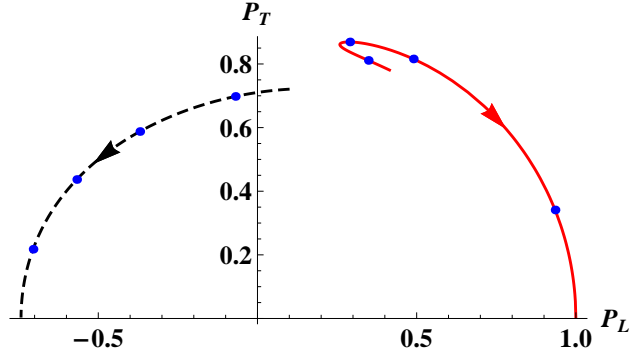}
\caption{The $q^2$ dependence of the orientation and the length of the polarization vector for the $B\to D^\ast$  (dashed line) and $B\to D$ (solid line) transitions. The arrows show the direction of increasing $q^2$. The dots on the dashed line stand for $q^2=4$, $6$, $8$, and $10~\text{GeV}^2$. The dots on the solid line stand for $q^2=4$, $8$, $10$, and $11.5~\text{GeV}^2$.}
\label{fig:apex}
\end{figure}

The $q^2$ dependence of the angle $\theta_{OD}$ is obviously related to the correlation between the longitudinal and transverse polarization components, or in other words, the orientation and the length of the polarization vector. In Fig.~\ref{fig:apex} we show how the apex of the polarization vector moves in the $(P_L,P_T)$ plane when $q^2$ increases from threshold $q^2=m_\tau^2$ to the zero-recoil points $q^2=(m_{\bar{B}^0}-m_{D^{(*)}})^2$. The apexes move within the unit circle since $|\vec{P}|\leq 1$. Both trajectories start off at threshold and end up at the zero-recoil points. As $q^2$ increases, the polarization vector of the $\tau^-$ turns into the direction of its three-momentum (for the $B\to D$ transition) or opposite to it (for the $B\to D^\ast$ transition). Both transverse
polarization components vanish at zero recoil as follows from the helicity
analysis in Sec.~\ref{sec:formalism}.
It is interesting to note that in the case of the $B\to D^\ast$ transition the dots are approximately equally spaced on the trajectory, which indicates a moderate rotation of the polarization vector when $q^2$ increases. In contrast, the polarization vector in the case of the $B\to D$ transition rotates quite fast for $q^2 \gtrsim 10~\text{GeV}^2$. These behaviors are also reflected in the $q^2$ dependence of the angle $\theta_{OD}$ shown in Fig.~\ref{fig:OD_angle}. The average values of the polar angle $\theta_{OD}$
  read $<\theta_{OD}>=130\degree$ for $B \to D^\ast$ and
  $<\theta_{OD}>=64\degree$ for $B \to D$.
\section{Effective operators and helicity amplitudes}
\label{sec:formalism}
Assuming that all neutrinos are left-handed and that NP effects only influence
leptons of the third generation, the effective Hamiltonian for the quark-level
transition $b \to c \tau^- \bar{\nu}_{\tau}$ is given by
\bea
{\mathcal H}_{eff} &=&\frac{4G_F}{\sqrt{2}} V_{cb}\left[\mathcal{O}_{V_L}+
\sum\limits_{X=S_L,S_R,V_L,V_R,T_L} X\mathcal{O}_{X}\right],
\label{eq:Heff}
\ena
where the four-Fermi operators $\mathcal{O}_{X}$ are defined as
\bea
\mathcal{O}_{V_L} &=& 
\left(\bar{c}\gamma^{\mu}P_Lb\right)\left(\bar{\tau}\gamma_{\mu}P_L\nu_{\tau}
\right),
\nn
\mathcal{O}_{V_R} &=&
\left(\bar{c}\gamma^{\mu}P_Rb\right)
\left(\bar{\tau}\gamma_{\mu}P_L\nu_{\tau}\right),
\nn
\mathcal{O}_{S_L} &=& \left(\bar{c}P_Lb\right)\left(\bar{\tau}P_L\nu_{\tau}\right),
\\
\mathcal{O}_{S_R} &=& \left(\bar{c}P_Rb\right)\left(\bar{\tau}P_L\nu_{\tau}\right),
\nn
\mathcal{O}_{T_L} &=& \left(\bar{c}\sigma^{\mu\nu}P_Lb\right)
\left(\bar{\tau}\sigma_{\mu\nu}P_L\nu_{\tau}\right),\nonumber
\label{eq:operators}
\ena
and $X$'s are the NP complex Wilson coefficients which are equal to zero in the SM.

The invariant form factors describing the hadronic transitions $\bar{B}^0 \to D$
and $\bar{B}^0 \to D^\ast$ are defined as follows:
\bea
\label{ff}
\langle D(p_2)
|\bar{c} \gamma^\mu b
| \bar{B}^0(p_1) \rangle
&=& F_+(q^2) P^\mu + F_-(q^2) q^\mu,\nn
\langle D(p_2)
|\bar{c}b
| \bar{B}^0(p_1) \rangle &=& (m_1+m_2)F^S(q^2),\nn
\langle D(p_2)|\bar{c}\sigma^{\mu\nu}(1-\gamma^5)b|\bar{B}^0(p_1)\rangle 
&=&\frac{iF^T(q^2)}{m_1+m_2}\left(P^\mu q^\nu - P^\nu q^\mu 
+i \varepsilon^{\mu\nu Pq}\right),\nn
\langle D^\ast(p_2)
|\bar{c} \gamma^\mu(1\mp\gamma^5)b
| \bar{B}^0(p_1) \rangle
&=& \frac{\epsilon^{\dagger}_{2\alpha}}{m_1+m_2}
\Big[ \mp g^{\mu\alpha}PqA_0(q^2) \pm P^{\mu}P^{\alpha}A_+(q^2)\\
&&\pm q^{\mu}P^\alpha A_-(q^2) 
+ i\varepsilon^{\mu\alpha P q}V(q^2)\Big],
\nn
\langle D^\ast(p_2)
|\bar{c}\gamma^5 b
| \bar{B}^0(p_1) \rangle &=& \epsilon^\dagger_{2\alpha}P^\alpha G^S(q^2),
\nn
\langle D^\ast(p_2)|\bar{c}\sigma^{\mu\nu}(1-\gamma^5)b|\bar{B}^0(p_1)\rangle
&=&-i\epsilon^\dagger_{2\alpha}\Big[
\left(P^\mu g^{\nu\alpha} - P^\nu g^{\mu\alpha} 
+i \varepsilon^{P\mu\nu\alpha}\right)G_1^T(q^2)\nn
&&+\left(q^\mu g^{\nu\alpha} - q^\nu g^{\mu\alpha}
+i \varepsilon^{q\mu\nu\alpha}\right)G_2^T(q^2)\nn
&&+\left(P^\mu q^\nu - P^\nu q^\mu 
+ i\varepsilon^{Pq\mu\nu}\right)P^\alpha\frac{G_0^T(q^2)}{(m_1+m_2)^2}
\Big],\nonumber
\ena
where $P=p_1+p_2$, $q=p_1-p_2$, and $\epsilon_2$ is the polarization vector
of the $D^\ast$ meson which satisfies the condition $\epsilon_2^\dagger\cdot p_2=0$.
The particles are on their mass shells: $p_1^2=m_1^2=m_{\bar{B}^0}^2$ and
 $p_2^2=m_2^2=m_{D^{(\ast)}}^2$.

Using the helicity technique first described 
in Refs.~\cite{Korner:1987kd,Korner:1989ve,Korner:1989qb} and
further discussed in our recent papers~\cite{Ivanov:2015tru,Ivanov:2016qtw} one obtains
the ratio of branching fractions $R_{D^{(*)}}(q^2)$ as follows:
\be
R_{D^{(*)}}(q^2)
=\left(\frac{q^2-m_\tau^2}{q^2-m_\mu^2}\right)^2\frac{{\cal H}_{tot}^{D^{(\ast)}}}{\sum\limits_{n}|H_{n}|^2+\delta_\mu \big(\sum\limits_{n}|H_{n}|^2+3|H_{t}|^2\big)},
\en
where
\bea
{\cal H}_{tot}^{D}
&=& 
|1+g_V|^2\left[|H_0|^2+\delta_\tau(|H_0|^2+3|H_t|^2) \right]+\frac{3}{2}|g_S|^2 |H_P^S|^2\nn
&&+ 3\sqrt{2\delta_\tau} {\rm Re}g_S H_P^S H_t
+8|T_L|^2 ( 1+4\delta_\tau) |H_T|^2
+12\sqrt{2\delta_\tau} {\rm Re}T_L H_0 H_T,
\label{eq:HtotD}
\ena
\bea
{\cal H}_{tot}^{D^\ast}
&=&(|1+V_L|^2+|V_R|^2)\Big[\sum\limits_{n}|H_{n}|^2+\delta_\tau \Big(\sum\limits_{n}|H_{n}|^2+3|H_{t}|^2\Big) \Big]+\frac{3}{2}|g_P|^2|H^S_V|^2\nn
&&-2 {\rm Re}V_R\big[(1+\delta_\tau) (|H_{0}|^2+2H_{+}H_{-})+3\delta_\tau |H_{t}|^2 \big]
-3\sqrt{2\delta_\tau} {\rm Re}g_P H^S_V H_{t}\nn
&&+8|T_L|^2 (1+4\delta_\tau)\sum\limits_{n}|H_T^n|^2
-12\sqrt{2\delta_\tau} {\rm Re}T_L\sum\limits_{n} H_{n}H_T^n.
\label{eq:HtotDv}
\ena 
Here, $\delta_\ell = m^2_\ell/2q^2$ is the helicity flip factor, $g_V\equiv V_L+V_R$, $g_S\equiv S_L+S_R$, $g_P\equiv S_L-S_R$, and the index $n$ runs through $(0,+,-)$. The definition of the hadronic helicity amplitudes in terms of the invariant form factors are presented in the Appendix. The expressions for ${\cal H}_{tot}^{D^{(\ast)}}$ in Eqs.~(\ref{eq:HtotD}) and~(\ref{eq:HtotDv}) agree with the results of Ref.~\cite{Sakaki:2013bfa}. Note that in this paper we do not consider interference terms between different NP operators since we assume the dominance of only one NP operator besides the SM contribution. 

In the remaining part of this section we provide the formulae for the
polarization components of the $\tau^-$ including NP contributions. Starting from the definition given in Eq.~(\ref{eq:poldef}) one easily obtains the differential decay rate for a given spin projection in a given direction by using the Dirac projection operators, which results in the replacement of 
\be
\slp_\tau+m_\tau \rightarrow 
\frac{1}{2} (\slp_\tau+m_\tau)(1+\gamma_5\sls_i)
\en
in the relevant traces. The $W^-$ rest frame polarization vectors $s_i^\mu$ are
given by~\cite{Faessler:2002ut, Gutsche:2013oea, Gutsche:2015mxa}
\bea
s_L^\mu&=&\frac{1}{m_\tau}(|\vec{p}_\tau|,E_\tau\sin\theta_\tau,0,E_\tau\cos\theta_\tau),\nn
s_T^\mu&=&(0,\cos\theta_\tau,0,-\sin\theta_\tau),\nn
s_N^\mu&=&(0,0,1,0).
\ena
The longitudinal polarization reads
\bea
\label{eq:PL}
P_L^D(q^2)&=&\frac{1}{{\cal H}_{tot}^D}\Big\lbrace
-|1+g_V|^2\big[|H_0|^2-\delta_\tau(|H_0|^2+3|H_t|^2)\big]+3\sqrt{2\delta_\tau}{\rm Re}g_S H_P^S H_t\nn
&&+\frac{3}{2}|g_S|^2|H_P^S|^2
+8|T_L|^2(1-4\delta_\tau)|H_T|^2-4\sqrt{2\delta_\tau}{\rm Re}T_L H_0 H_T
\Big\rbrace,\nn
P_L^{D^\ast}(q^2)&=&\frac{1}{{\cal H}_{tot}^{D^\ast}}\Big\lbrace
(|1+V_L|^2+|V_R|^2)\big[-\sum\limits_{n}|H_{n}|^2+\delta_\tau(\sum\limits_{n}|H_{n}|^2+3|H_{t}|^2)\big]\nn
&&-2{\rm Re}V_R\big[(1-\delta_\tau)(-|H_{0}|^2+2H_{+}H_{-})+3\delta_\tau |H_{t}|^2\big]-3\sqrt{2\delta_\tau}{\rm Re}g_P H_V^S H_{t}\nn
&&+\frac{3}{2}|g_P|^2|H_V^S|^2+8|T_L|^2(1-4\delta_\tau)\sum\limits_{n}|H_T^n|^2+4\sqrt{2\delta_\tau}{\rm Re}T_L\sum\limits_{n}H_{n}H_T^n
\Big\rbrace
.
\ena
We emphasize that the longitudinal polarization of the $\tau^-$ is defined in
the $W^-$ rest frame with $\vec p_\tau$ defining the longitudinal direction, and
not in the rest frame of the parent $\bar{B}^0$ meson.

Similarly, the transverse polarization is given by
\bea
\label{eq:PT}
P_T^D(q^2)&=&\frac{3\pi\sqrt{\delta_\tau}}{2\sqrt{2}{\cal H}_{tot}^D}\Big\lbrace
|1+g_V|^2 H_0 H_t
+\frac{{\rm Re}g_S}{\sqrt{2\delta_\tau}} H_P^S H_0+4\sqrt{2\delta_\tau}{\rm Re}T_L H_t H_T
\Big\rbrace
,\nn
P_T^{D^\ast}(q^2)&=&\frac{3\pi\sqrt{\delta_\tau}}{4\sqrt{2}{\cal H}_{tot}^{D^\ast}}\Big\lbrace
(|1+V_L|^2-|V_R|^2)(|H_{-}|^2-|H_{+}|^2)+2(|1+V_L|^2+|V_R|^2)H_{t}H_{0}\nn
&&-\frac{2{\rm Re}g_P}{\sqrt{2\delta\tau}}H_V^SH_{0}-4{\rm Re}V_RH_{t}H_{0}+16|T_L|^2(|H_T^-|^2-|H_T^+|^2)\nn
&&+4{\rm Re}T_L\Big[\frac{1+2\delta_\tau}{\sqrt{2\delta_\tau}}(H_{+}H_T^+-H_{-}H_T^-)-2\sqrt{2\delta_\tau}H_{t}H_T^0\Big]
\Big\rbrace
.
\ena
As can be seen directly from Eq.~(\ref{eq:PT}), the transverse polarization
vanishes in the zero lepton mass limit $m_\ell=0$ due to the overall factor $\sqrt{\delta_\ell}=m_\ell/\sqrt{2q^2}$. Physically this comes about since the lepton is $100\%$ longitudinally
polarized for $m_\ell=0$ and thus there is no room for a transverse
polarization.  It is the lepton mass that brings in the transverse polarization which, in fact, is quite large in the
case of the $\tau^-$. In the SM the transverse polarization
can be seen to vanish at zero recoil as a result of the zero-recoil relations $H_t=0$ and $H_\pm=H_0$  (see the Appendix).

The normal polarization is zero in the SM because we take the form factors and
thereby the helicity amplitudes to be real. In the presence of NP
CP-violating complex Wilson
coefficients, they obtain nonzero contributions from the imaginary part of the coefficients as can be seen in Eq.~(\ref{eq:PN}). Both $P_N^D$ and $P_N^{D^\ast}$ are sensitive to the tensor and scalar operators. The normal polarization reads
\bea
\label{eq:PN}
P_N^D(q^2)&=&\frac{3\pi}{2{\cal H}_{tot}^D}\Big[
-{\rm Im}g_S H_P^S H_0+8\delta_\tau{\rm Im}T_L H_t H_T
\Big],\nn
P_N^{D^\ast}(q^2)&=&\frac{3\pi}{4{\cal H}_{tot}^{D^\ast}}\Big\lbrace
{\rm Im}g_P H_V^S H_{0}-2{\rm Im}T_L\big[(1-2\delta_\tau)(H_{+}H_T^+-H_{-}H_T^-)+4\delta\tau H_{t}H_T^0\big]
\Big\rbrace
.
\ena
\section{Numerical analysis}
\label{sec:analysis}
It is important to note that all the discussions and expressions that we have provided so far are model independent. Now, in order to make numerical predictions we use the form factors calculated in the covariant confined quark model (CCQM)~\cite{Ivanov:2016qtw} which has been developed in several previous papers by our group (see Refs.~\cite{Branz:2009cd,Ivanov:2011aa,Ivanov:2015woa} and references therein). One can also employ the form factors obtained from the heavy quark effective theory (HQET) with better controlled errors. However, in this section, we only aim at clarifying the role of the tau polarization in searching for NP; therefore, the use of our form factors is well suited. For example, the longitudinal polarizations calculated in our model assuming only the SM operator
are $<P^D_L>=0.33$ and $<P^{D^\ast}_L>=-0.50$, which are in very good agreement
with other results in the literature
$<P^D_L>=0.325\pm0.009$~\cite{Tanaka:2010se} and
$<P^{D^\ast}_L>=-0.497\pm0.013$~\cite{Tanaka:2012nw, Hirose:2016wfn}.
\subsection{Form factors in the CCQM}
\label{subsec:FF}
 As has been discussed in detail in Ref.~\cite{Ivanov:2016qtw} we calculate the
      current-induced $B \to D^{(\ast)}$ transitions from their one-loop quark
      diagrams. As a result
the various form factors in our model are represented by three-fold integrals
which are calculated by using \textsc{fortran} codes in the full kinematical
momentum 
transfer region $0\le q^2 \le q^2_{max}=(m_{\bar{B}^0}-m_{D^{(\ast)}})^2$. Our numerical results for the form factors are well represented
by a double-pole parametrization
\be
F(q^2)=\frac{F(0)}{1 - a s + b s^2}, \quad s=\frac{q^2}{m_1^2}. 
\label{eq:DPP}
\en 
The parameters of the  form factors for the $\bar{B}^0 \to D$ and $\bar{B}^0 \to D^\ast$ transitions are listed in Table~\ref{tab:ff-param}.
\begin{table}[ht]
\caption{Parameters of the dipole approximation in Eq.~(\ref{eq:DPP}) for  $\bar{B}^0 \to D^{(\ast)}$ form factors. Zero-recoil values of the form factors are also listed for comparison with the HQET.}
\begin{center}
\begin{tabular}{lccccccccccccc}
\hline\hline
\multicolumn{1}{c}{} &\multicolumn{8}{c}{$\bar{B}^0 \to D^\ast$} &\multicolumn{1}{c}{} 
                      &\multicolumn{4}{c}{$\bar{B}^0 \to D$} \\
\cline{2-9}\cline{11-14}
 & $ A_0 $ & $  A_+  $ & $  A_-  $ & $  V  $ 
 & $ G^S $ & $G_0^T$ & $G_1^T$ & $  G_2^T$ & {} & $F_+$ & $F_-$ & $F^S$ & $F^T$ 
 \\
\hline
$F(0)$ &  1.62 & 0.67  & $-0.77$ & 0.77 & $-0.50$ & $-0.073$ & 0.73 & $-0.37$ & {} &  0.79   & $-0.36$ &  0.80 & 0.77  
\\
$a$    &  0.34 & 0.87  &  0.89 & 0.90 & 0.87 & 1.23 & 0.90 & 0.88 & {} &  0.75   &  0.77 &  0.22 & 0.76  
\\
$b$    & $-0.16$ & 0.057 & 0.070 & 0.075 & 0.060 & 0.33 & 0.074 & 0.065 & {} &  0.039  & 0.046 & $-0.098$ & 0.043 
\\ 
$F(q^2_{\rm max})$ &  1.91 & 0.99  & $-1.15$ & 1.15 & $-0.74$ & $-0.13$ & 1.10 & $-0.55$ & {} &  1.14   & $-0.53$ &  0.89 & 1.11  
\\
$F^{HQET}(q^2_{\rm max})$ &  1.99 & 1.12  & $-1.12$ & 1.12 & $-0.62$ & 0 & 1.12 & $-0.50$ & {} &  1.14   & $-0.54$ &  0.88 & 1.14  
\\
\hline\hline
\end{tabular}
\label{tab:ff-param}
\end{center}
\end{table}
We also list the zero-recoil values of the form factors for comparison with the corresponding HQET results which can e.g. be found in Ref.~\cite{Ivanov:2016qtw}. The agreement between the two sets of
zero-recoil values is within $10 \%$. It is worth mentioning that we obtain a nonzero result for the form factor
$G_0^T$ at zero recoil, which is predicted to vanish in the HQET. 

\begin{figure}[htbp]
\begin{tabular}{cc}
\includegraphics[scale=0.5]{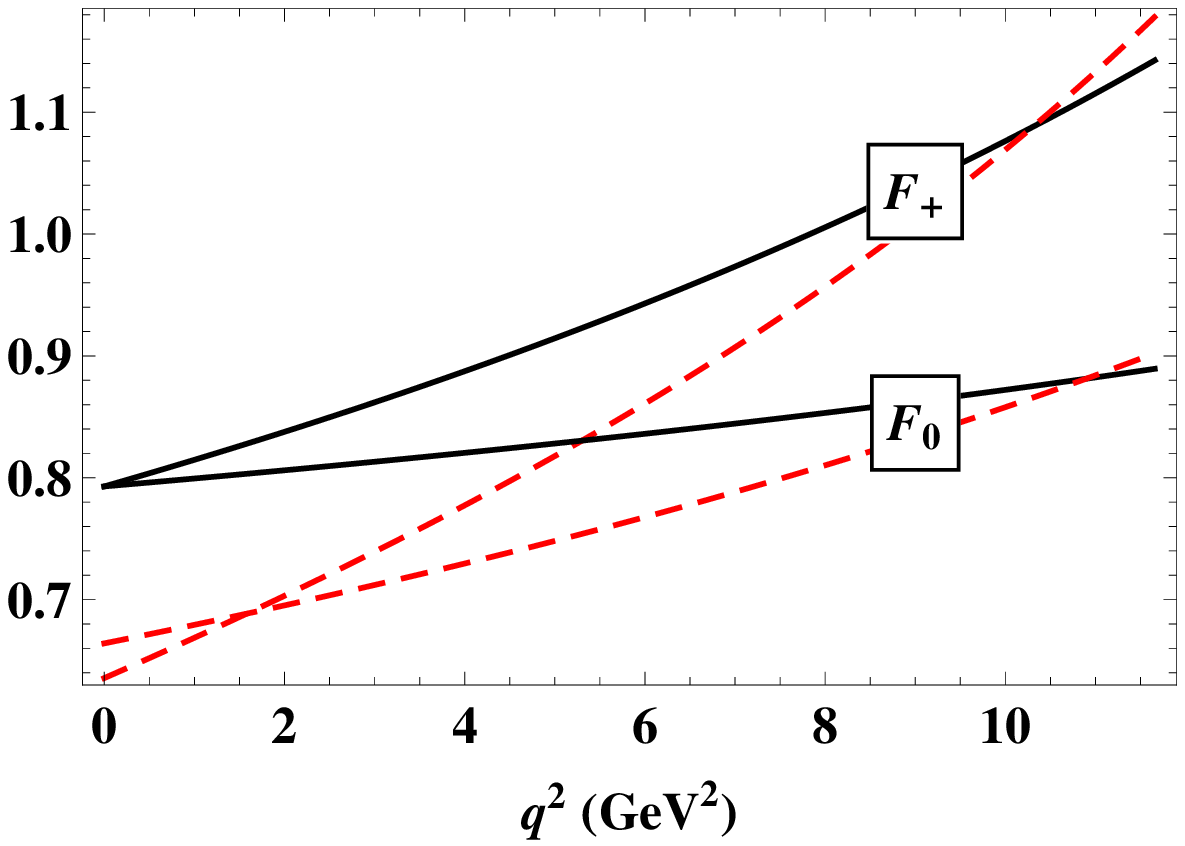}&
\includegraphics[scale=0.5]{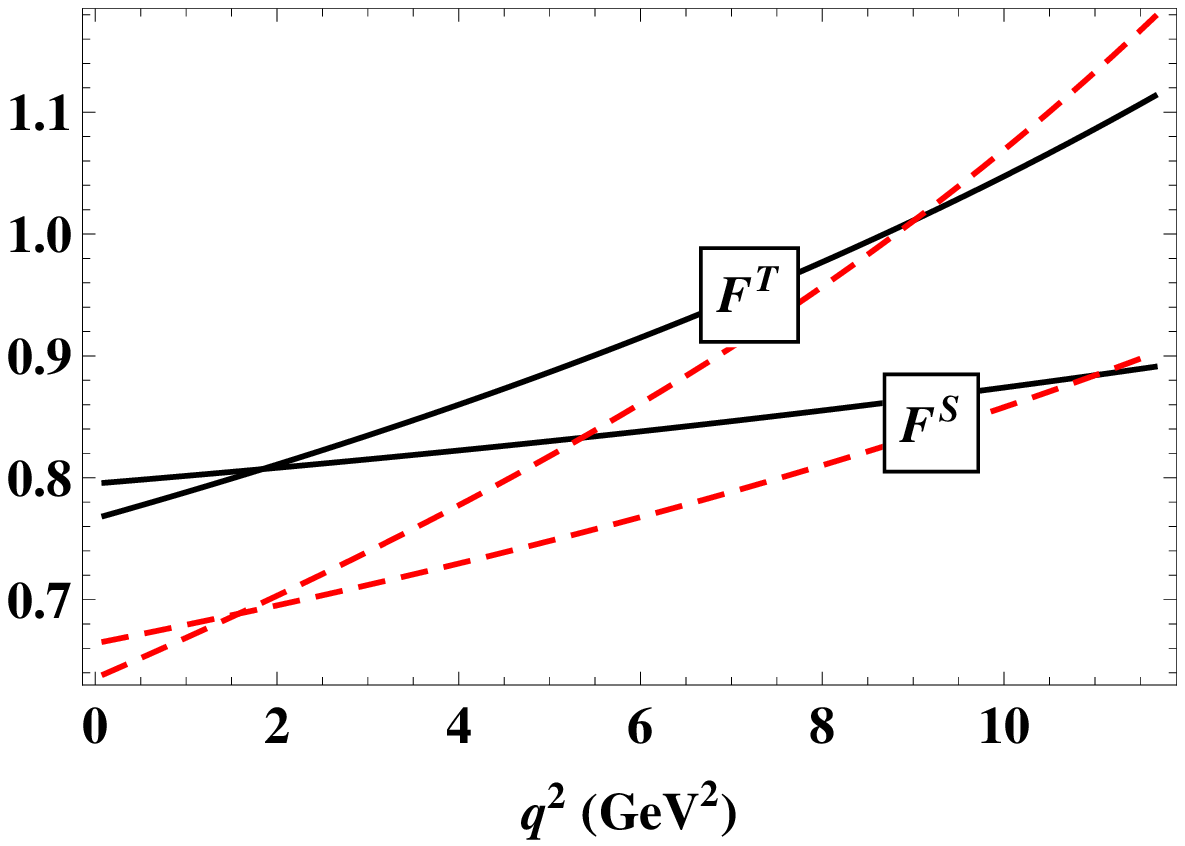}
\\
\includegraphics[scale=0.5]{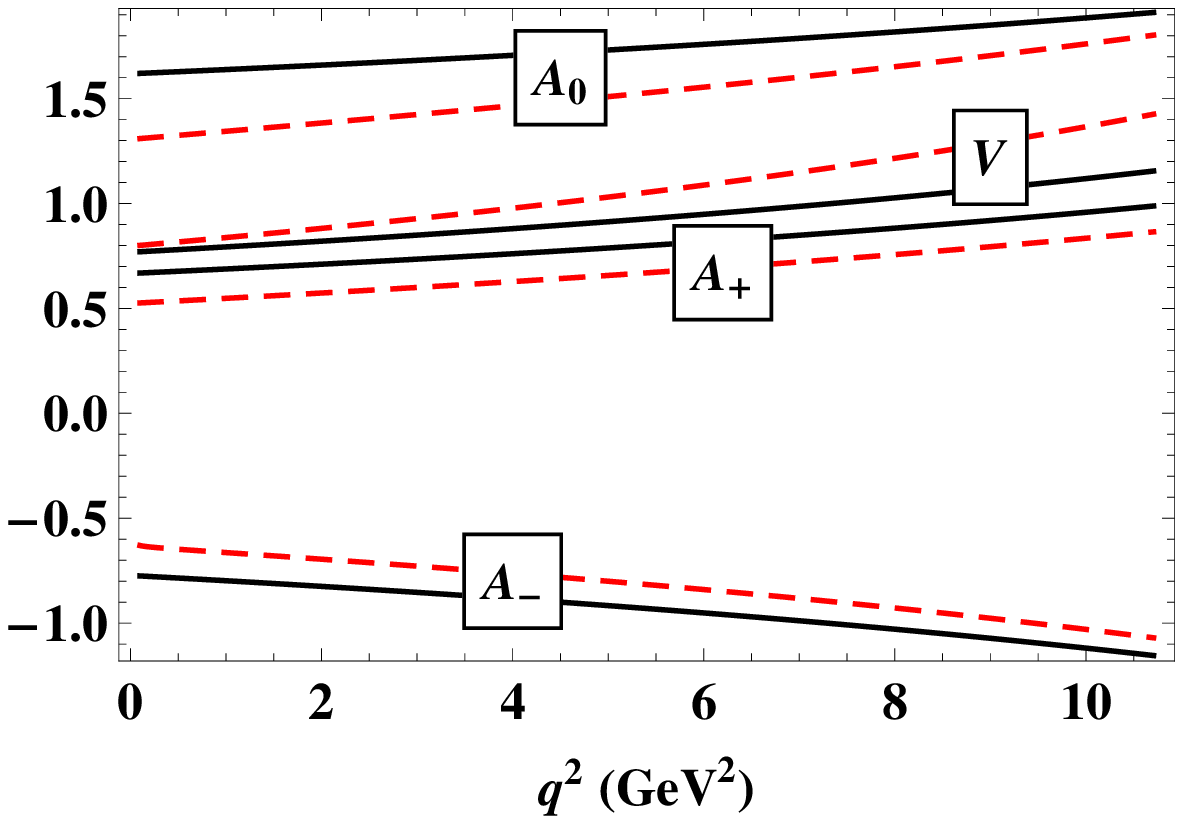}&
\includegraphics[scale=0.5]{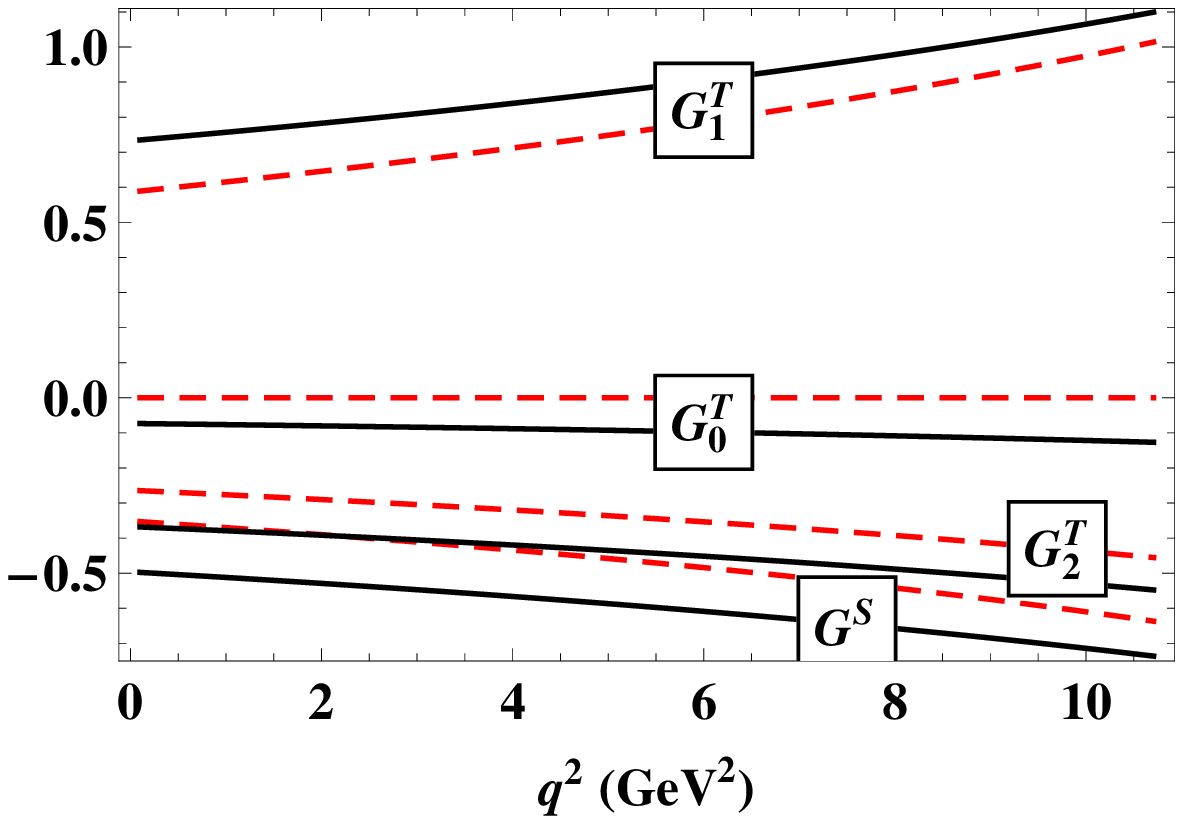}
\end{tabular}
\caption{Comparison of our form factors (solid lines) with the AKC form factors~\cite{Alonso:2016gym} (dashed lines) for the $\bar{B}^0\to D$ (upper panels) and $\bar{B}^0\to D^\ast$ (lower panels) transitions. Each CCQM form factor is labeled together with the corresponding AKC one by a box with their name put on both lines.}
\label{fig:FF}
\end{figure}

In Fig.~\ref{fig:FF}, we compare our form factors with the
Alonso-Kobach-Camalich (AKC) form factors calculated in
Ref.~\cite{Alonso:2016gym} where they have used theoretical input from the HQET,
lattice calculations, and equation of motion (EOM) relations. We rewrite the
AKC form factors in our notation using the relations between the two sets of form factors. The form factor $F_0(q^2)$ in Fig.~\ref{fig:FF} is given by
\be
F_0(q^2)=F_+(q^2)+\frac{q^2}{m_1^2-m_2^2}F_-(q^2).
\en
It is seen that our form factors share quite similar shapes with the corresponding AKC ones. 
The first plot in Fig.~\ref{fig:FF} shows that our form factors $F_+(q^2)$ and $F_0(q^2)$ (solid lines) satisfy the relation $F_0(0)=F_+(0)$ while the corresponding AKC form factors (dashed lines) are slightly different at $q^2=0$. This is due to the fact that in their paper~\cite{Alonso:2016gym}, the authors used different parametrizations for $F_+(q^2)$ and $F_0(q^2)$. More specifically, they used the
Caprini-Lellouch-Neubert parametrization for $F_+(q^2)$~\cite{Caprini:1997mu,Tanaka:2012nw}, but the Bourrely-Caprini-Lellouch parametrization for $F_0(q^2)$~\cite{Bourrely:2008za,Na:2015kha}. However, the difference $F_0(0)-F_+(0)\approx 0.03$ lies within the uncertainty of $F_+(q^2)$ at $q^2=0$, which reads $F_+(0)=0.664(34)$~\cite{Na:2015kha}.

We note that in Ref.~\cite{Ivanov:2015tru} the heavy quark limit (HQL) in our approach was explored in great detail  for the heavy-to-heavy $\bar{B}^0\to D^{(\ast)}$ transitions. In Ref.~\cite{Ivanov:2015tru} we also calculated the Isgur-Wise function and considered the near-recoil behavior of the form factors. A brief discussion of the subleading corrections to the HQL arising from finite quark masses can be found in Appendix~B of our paper~\cite{Ivanov:2016qtw}. Note that our form factors
do not satisfy the EOM relations since the $b$ and $c$ quarks in the
relevant propagators in the quark loop are off their mass shells.

Finally, we briefly discuss some error estimates within our model.
We fix our model parameters (the constituent quark masses, the infrared cutoff,
and the hadron size parameters) by minimizing the functional
$\chi^2 = \sum\limits_i\frac{(y_i^{\rm expt}-y_i^{\rm theor})^2}{\sigma^2_i}$
where $\sigma_i$ is the experimental  uncertainty.
If $\sigma$ is too small then we take its value of 10$\%$.
Moreover, we observed that the errors of the fitted parameters 
are of the order of  10$\%$.
Thus we estimate the model uncertainties to lie within 10$\%$.
\subsection{Experimental constraints}
\label{subsec:constraint}
\begin{figure}[htbp]
\begin{tabular}{cc}
\includegraphics[scale=0.35]{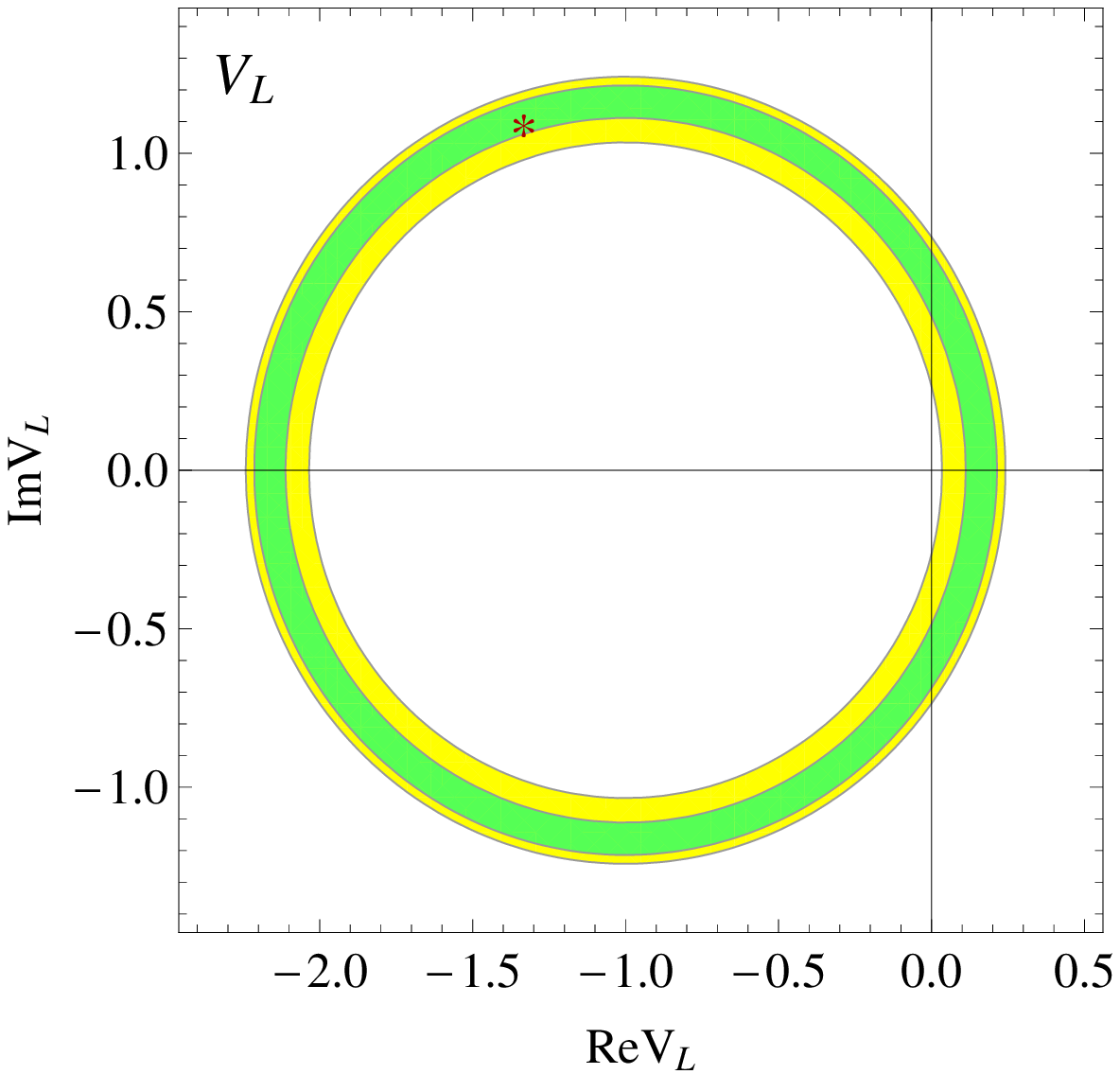}&
\includegraphics[scale=0.35]{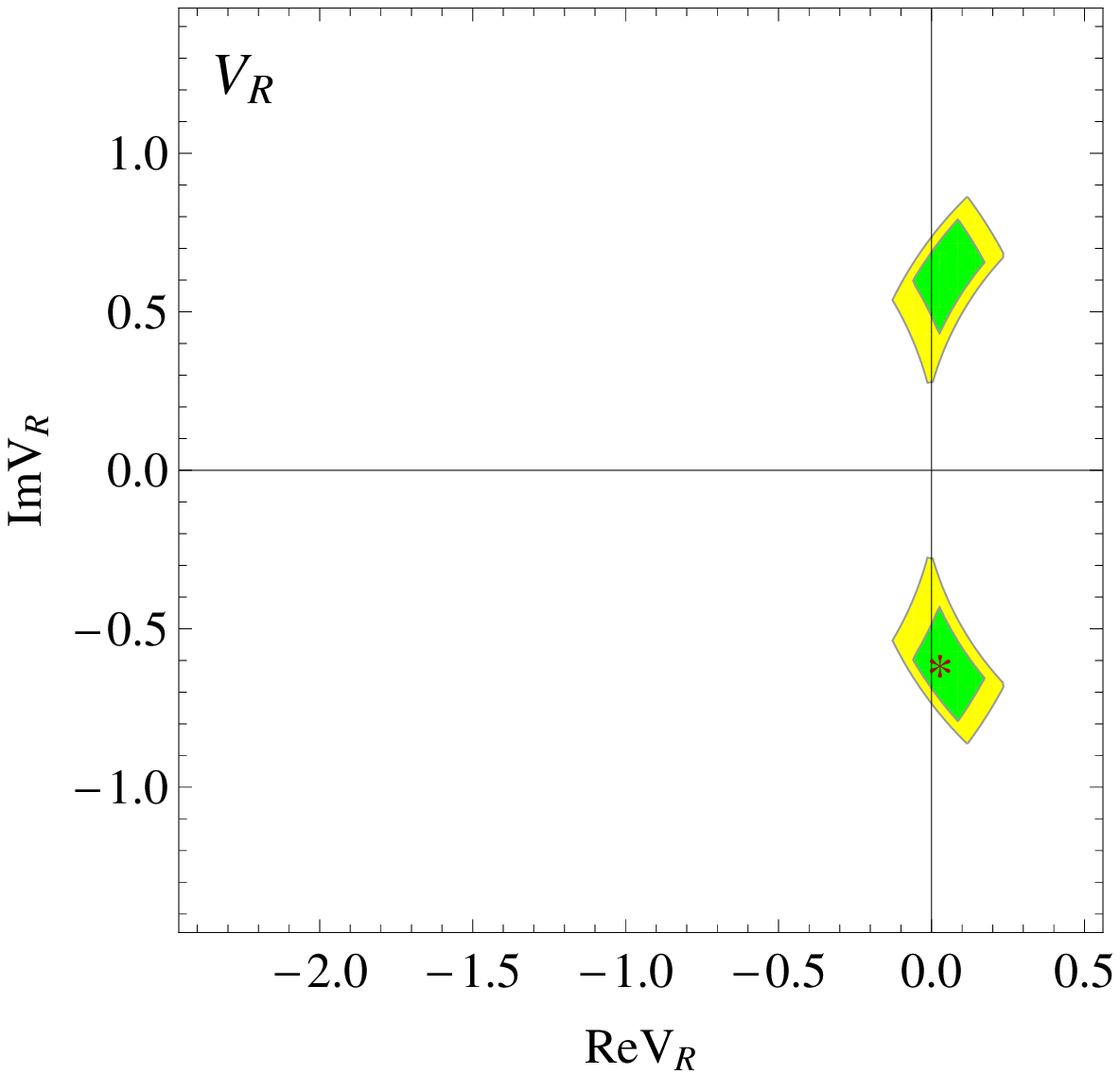}\\
\includegraphics[scale=0.35]{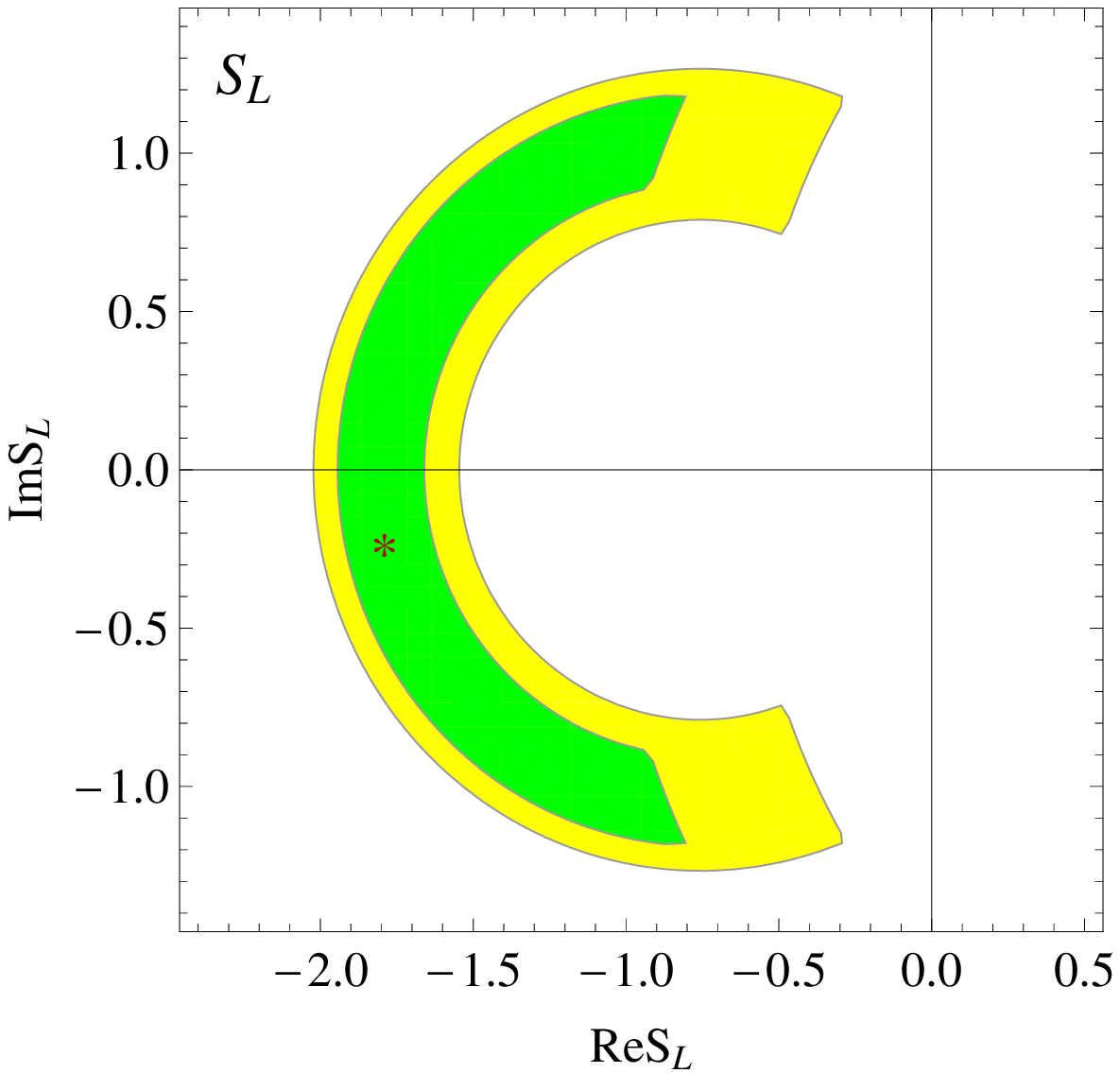}
& 
\includegraphics[scale=0.35]{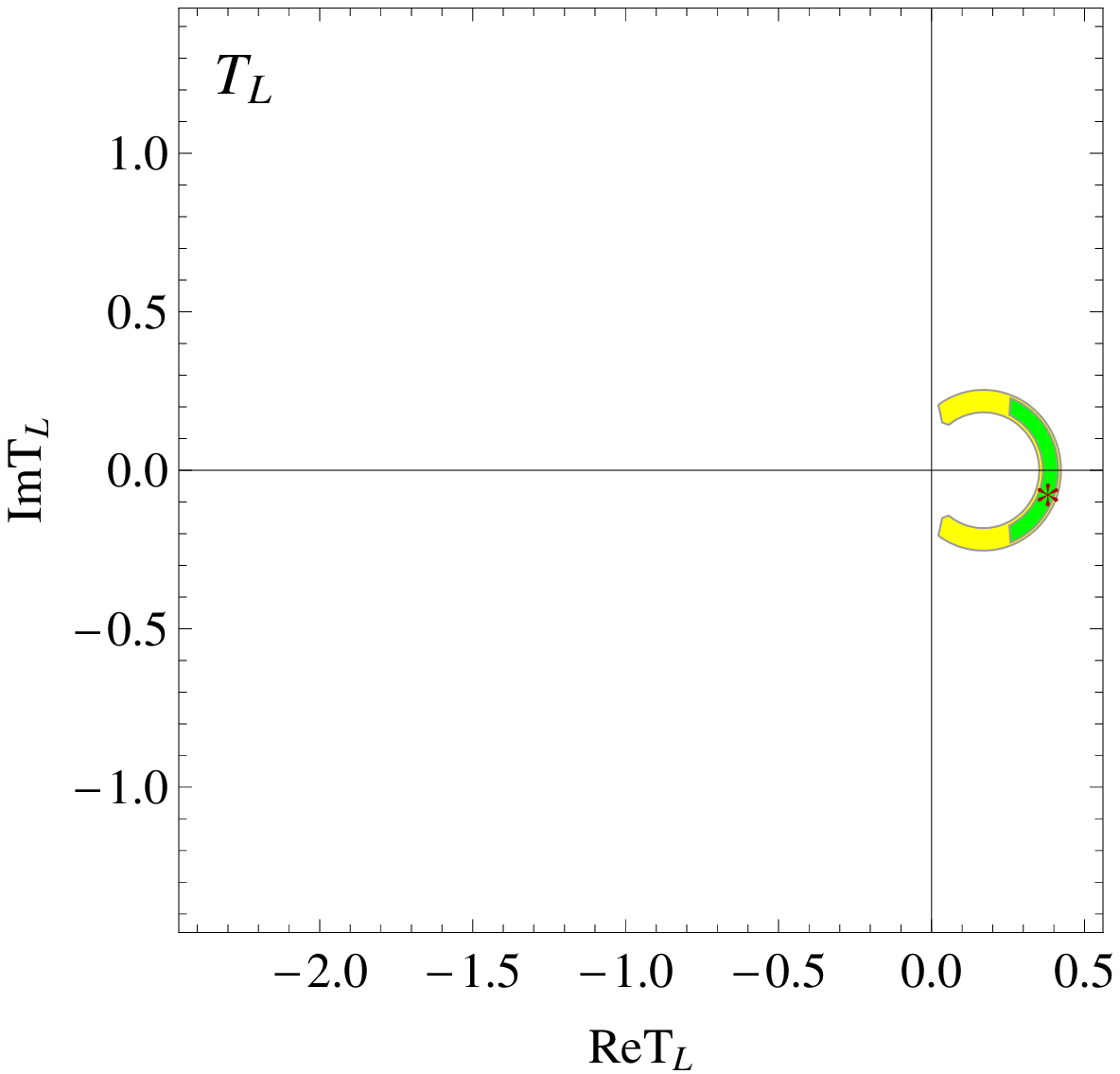}
\end{tabular}
\caption{Constraints on the Wilson coefficients $V_L$, $V_R$, $S_L$, and $T_L$ within $1\sigma$ (green, dark) and $2\sigma$ (yellow, light). No value of $S_R$ is allowed within $2\sigma$. The best-fit value in each case is denoted with the symbol $\ast$.}
\label{fig:constraint}
\end{figure}

Within the SM (without any NP operators) our model calculation yields
$R(D)=0.267$ and $R(D^\ast)=0.238$,
which are consistent with other SM predictions given in Refs.~\cite{Na:2015kha,Lattice:2015rga,Aoki:2016frl, Fajfer:2012vx} within $10\%$. Assuming the dominance of only one NP operator in Eq.~(\ref{eq:Heff}) at a time (besides the SM contribution), we compare the calculated ratios $R_{D^{(\ast)}}$ with the current experimental data $R_D = 0.406 \pm 0.050$ and $R_{D^\ast} = 0.311\pm 0.016$ given in Sec.~\ref{sec:intro} and obtain the allowed regions for the NP couplings as shown in Fig.~\ref{fig:constraint}. 
It is important to note that while determining these regions, we also take into account a theoretical error of $10\%$ for the ratios $R(D^{(\ast)})$. The operator $\mathcal{O}_{S_R}$ is excluded at $2\sigma$ and is not presented here. The operator $\mathcal{O}_{V_L}$ is not excluded, but it does not affect the polarizations in general and, therefore, will not be considered in what follows. In other words, only three NP operators $\mathcal{O}_{V_R}$, $\mathcal{O}_{S_L}$, and $\mathcal{O}_{T_L}$ can modify the polarizations. In each allowed region at $2\sigma$ we find the best-fit value for each NP coupling. The best-fit couplings read
\be
V_L =-1.33+i1.11,\quad V_R =0.03-i0.60,\quad
S_L =-1.79-i0.22, \quad T_L =0.38-i0.06,
\label{eq:bestfit}
\en
and are marked with an asterisk.
\subsection{Theoretical predictions}
\label{prediction}
\begin{figure}[htbp]
\begin{tabular}{lll}
\includegraphics[scale=0.4]{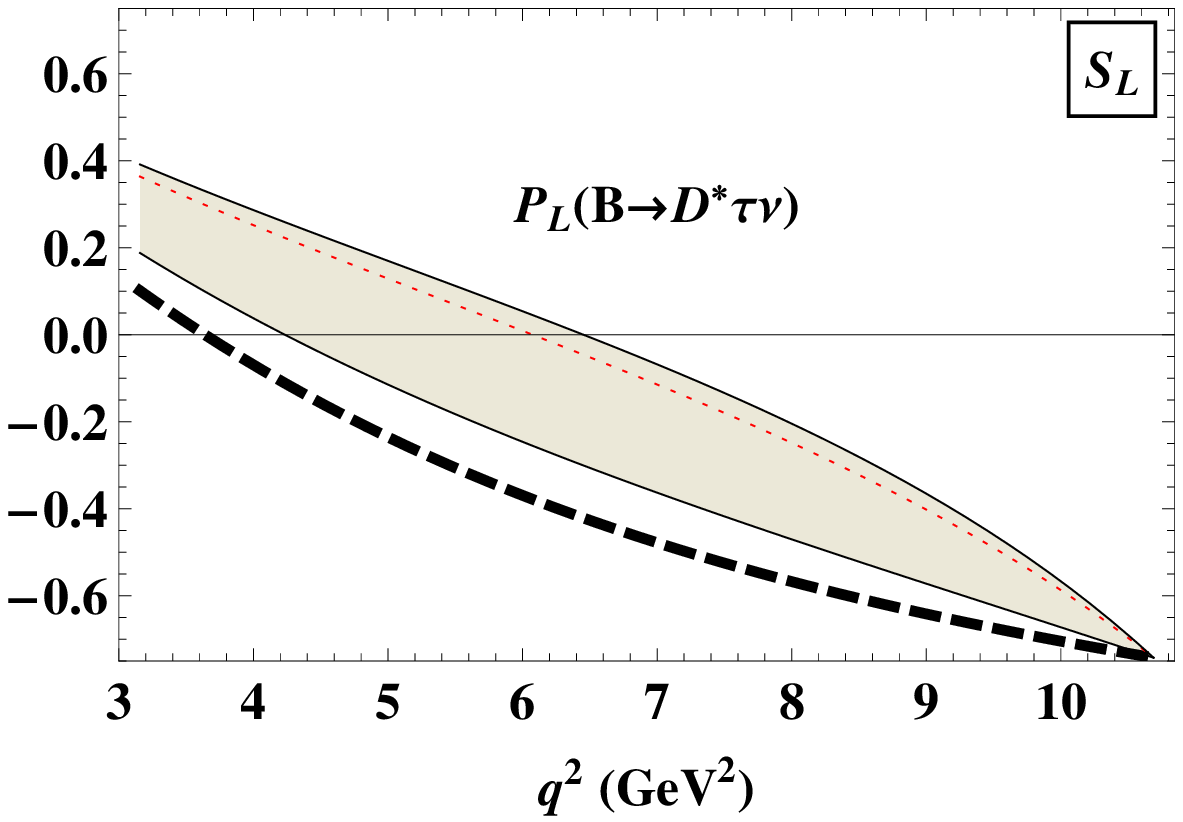}
& 
\includegraphics[scale=0.4]{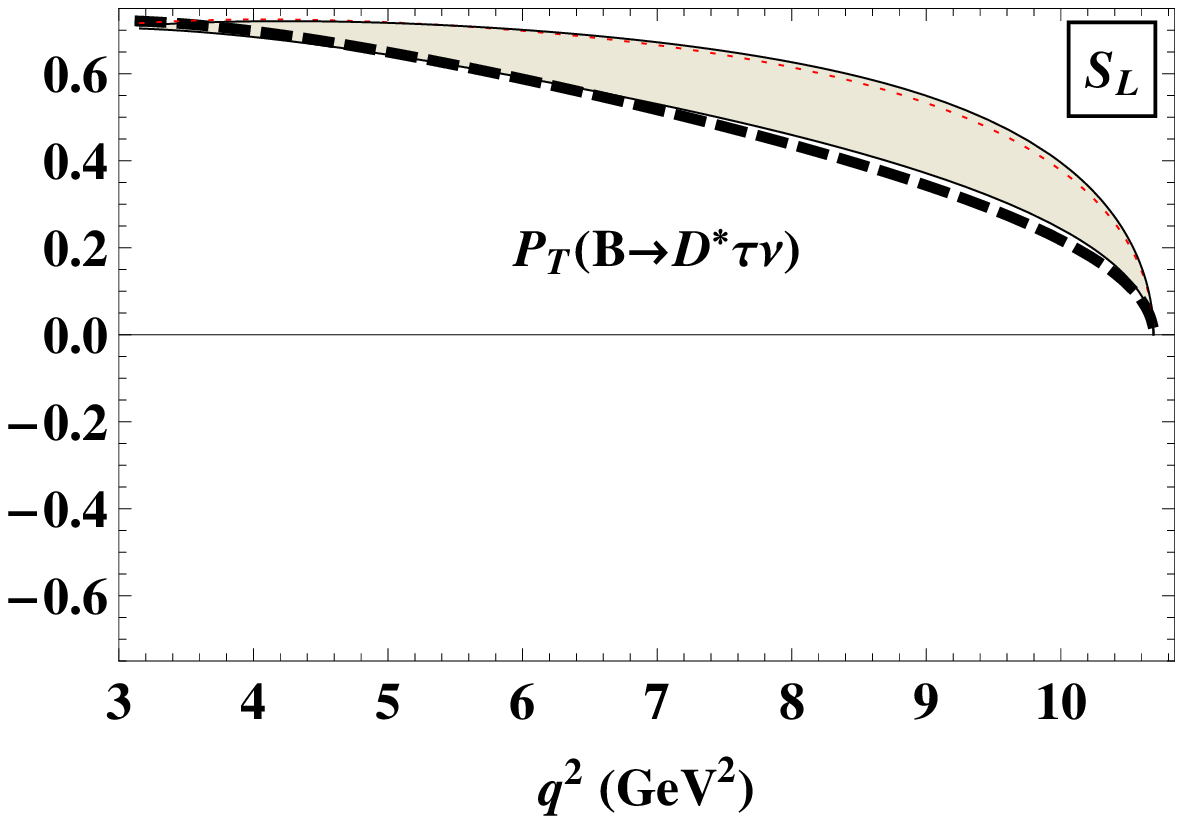}
&
\includegraphics[scale=0.4]{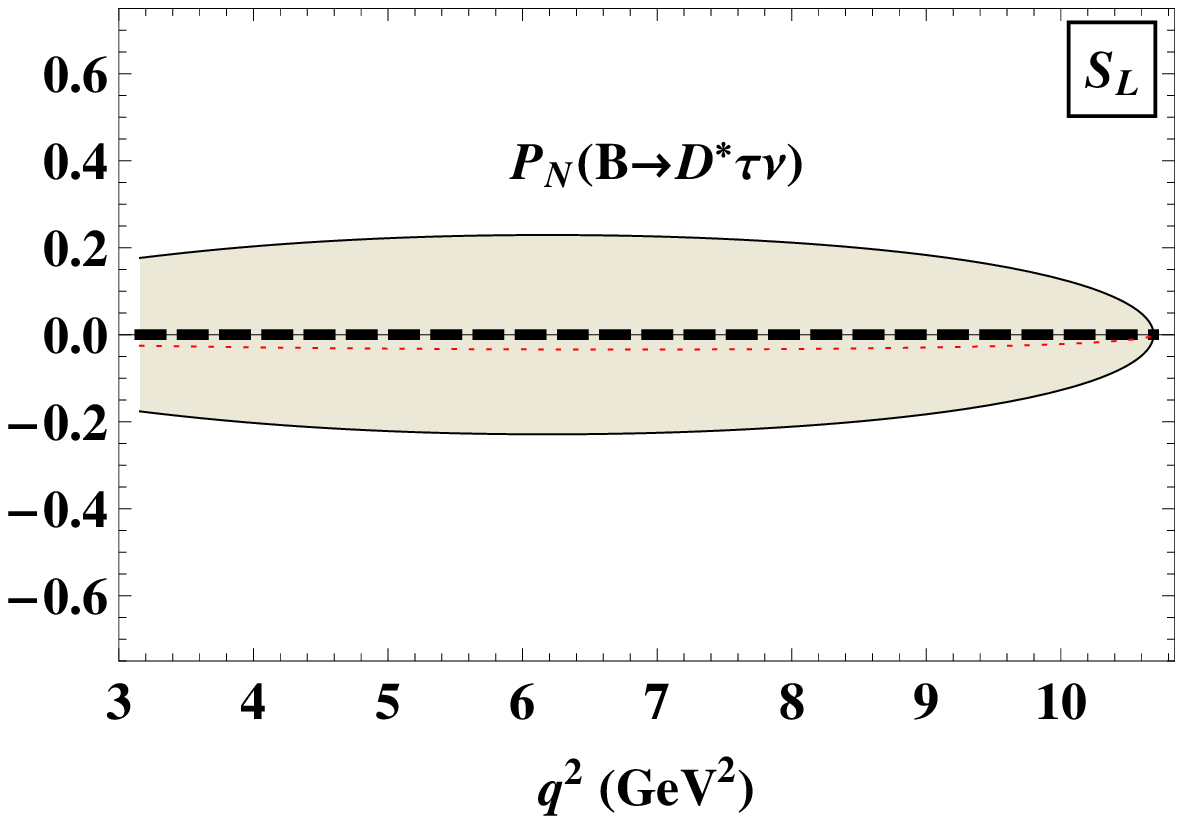}\\
\includegraphics[scale=0.4]{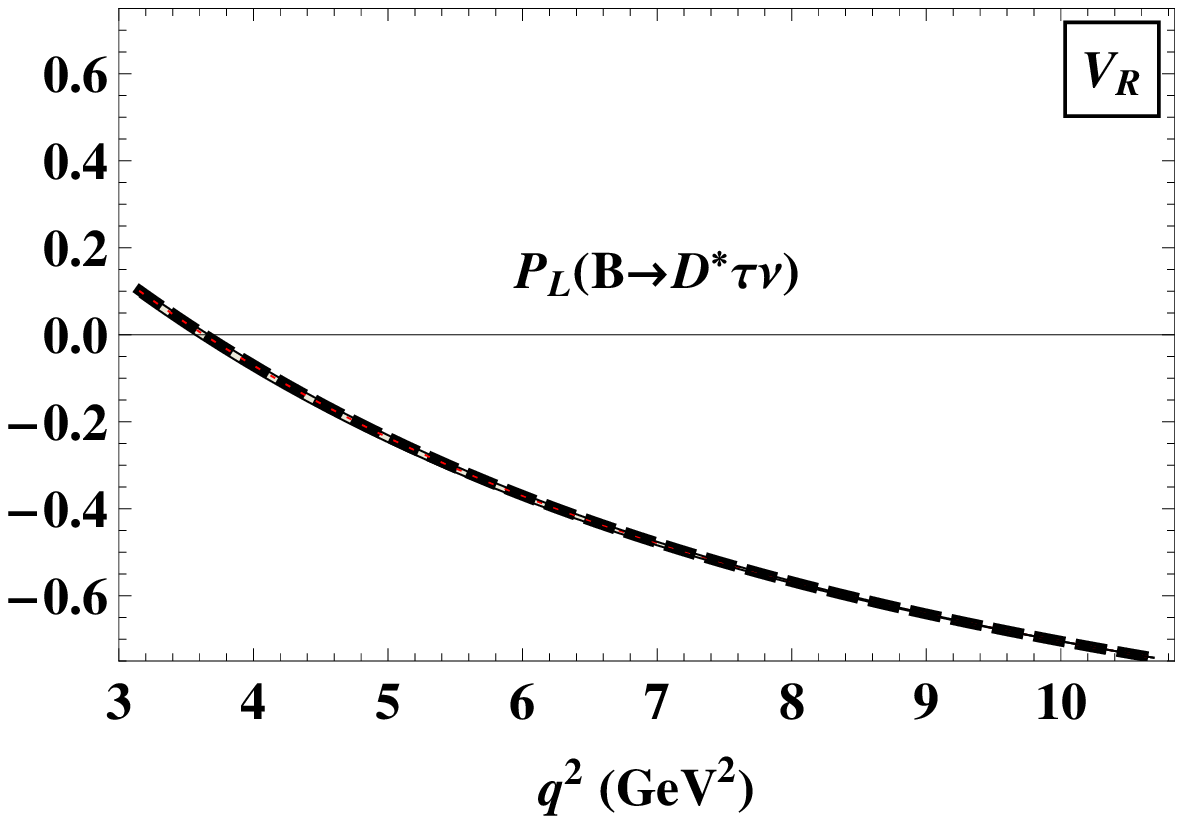}
& 
\includegraphics[scale=0.4]{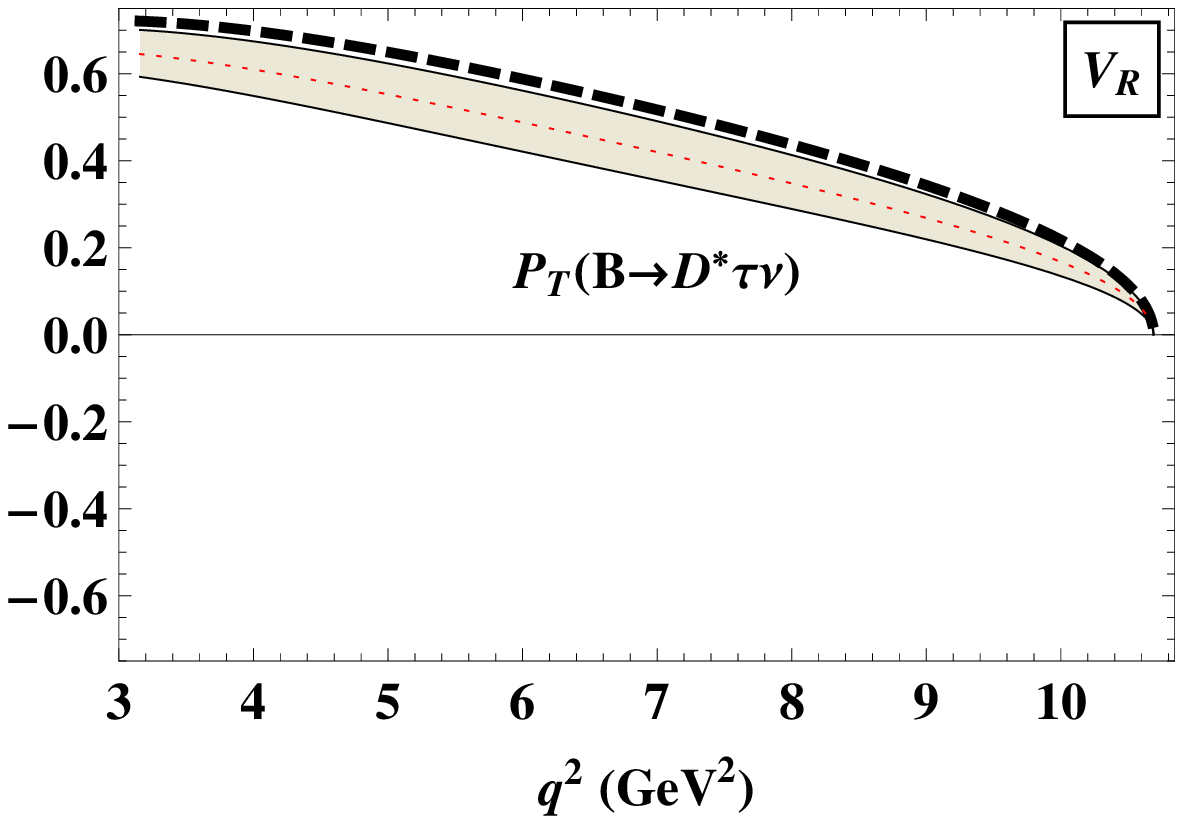}
&
\includegraphics[scale=0.4]{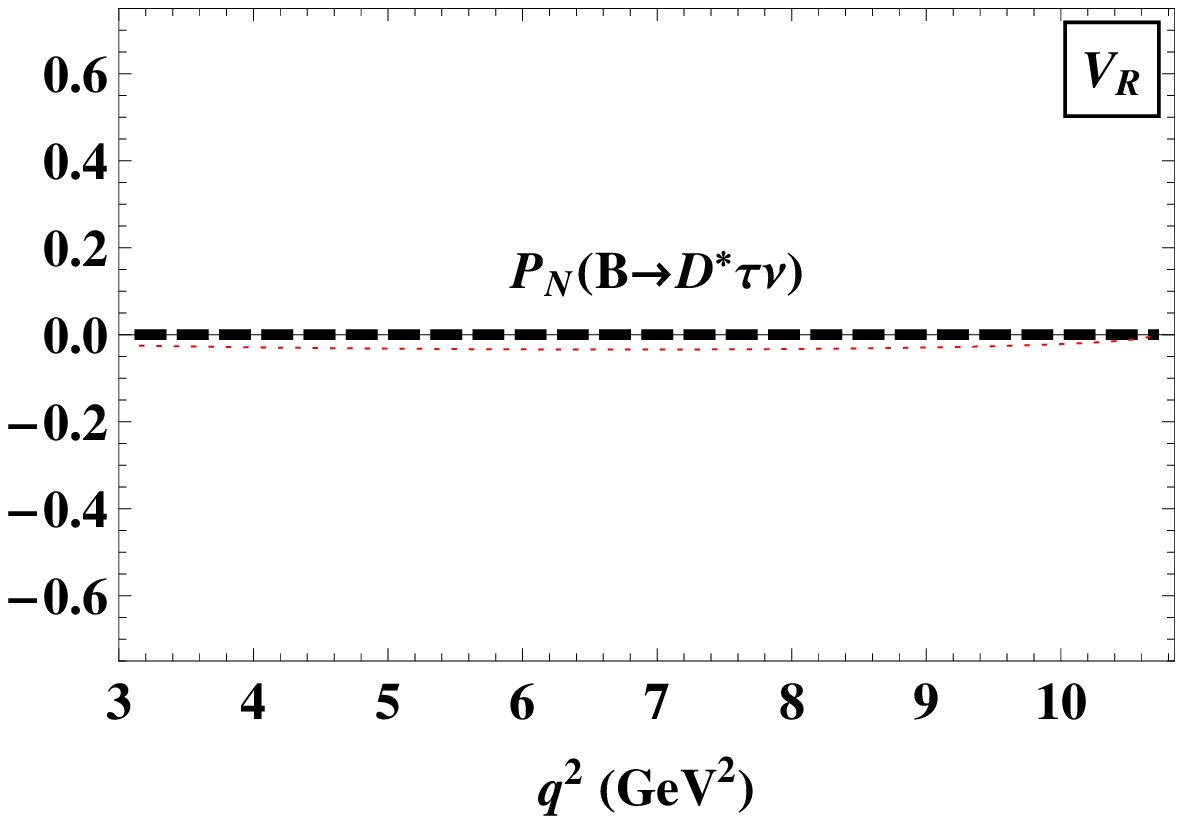}
\\
\includegraphics[scale=0.4]{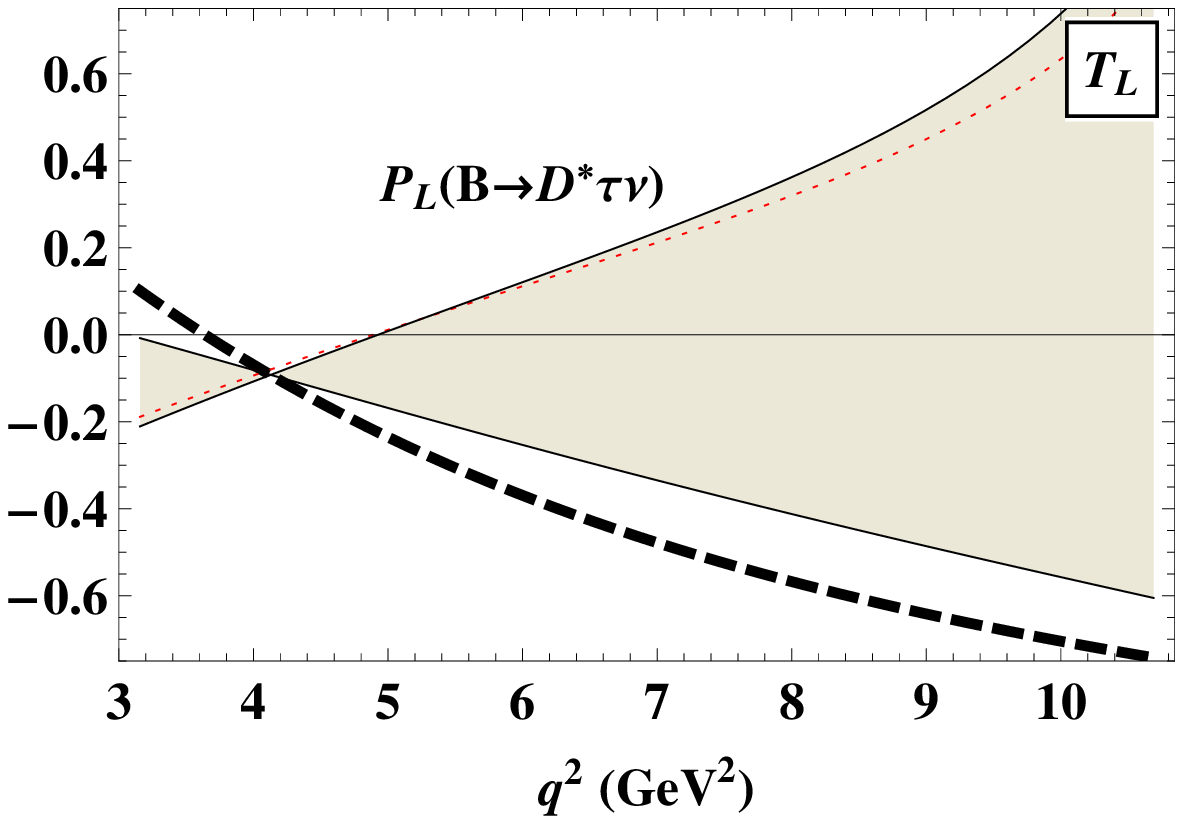}
& 
\includegraphics[scale=0.4]{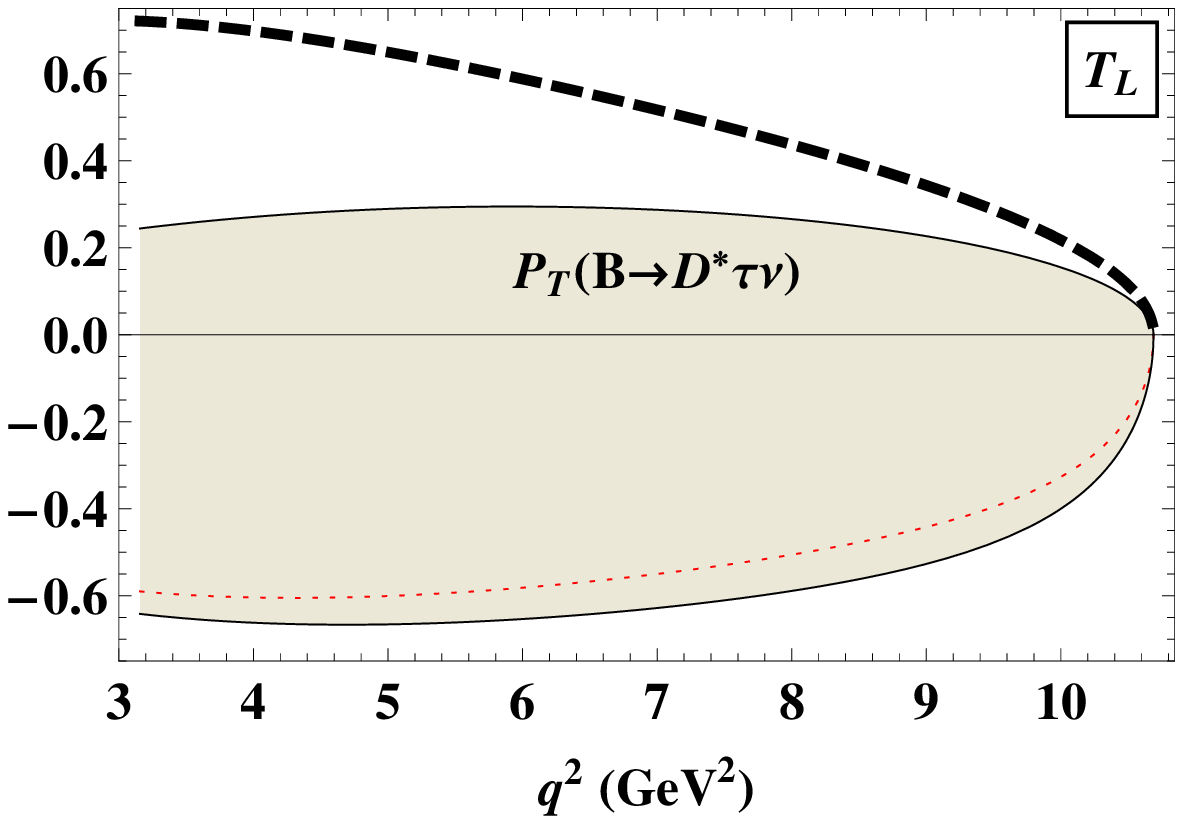}
& 
\includegraphics[scale=0.4]{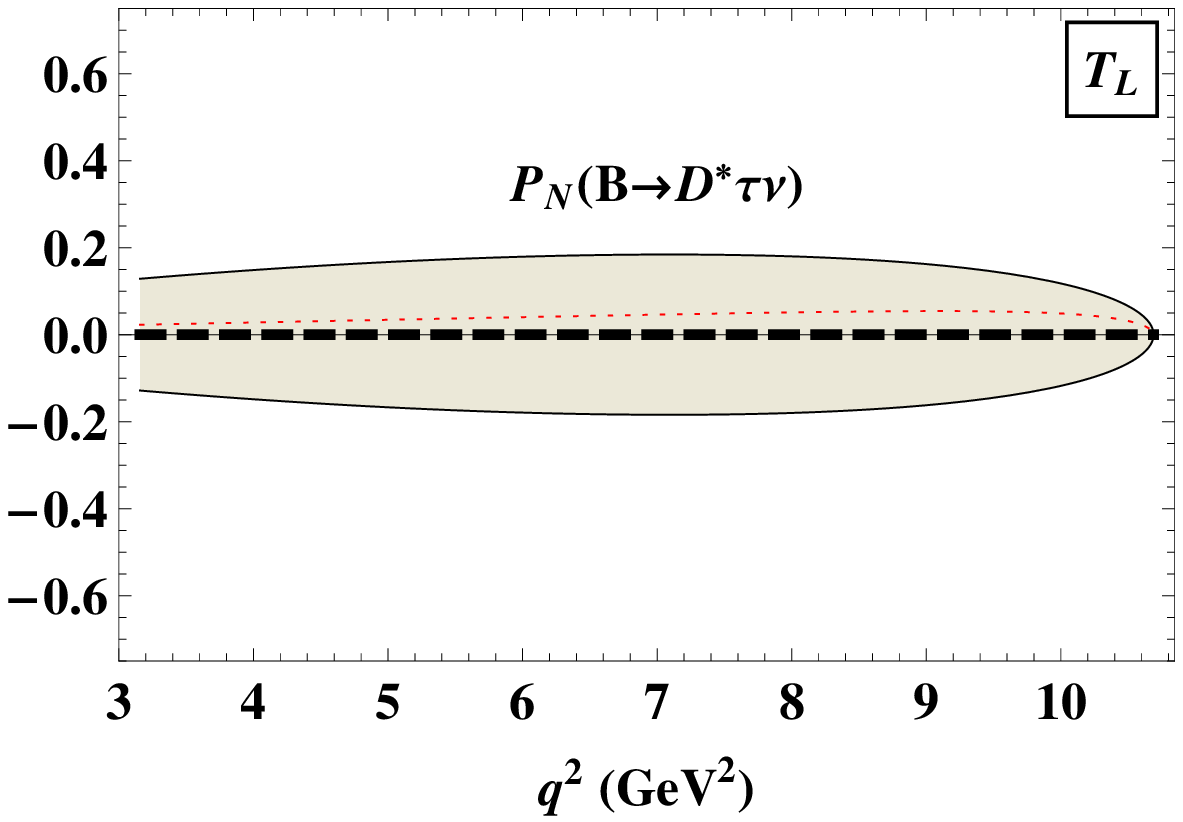}
\end{tabular}
\caption{Longitudinal (left), transverse (center), and normal (right) polarization of the $\tau^-$ in the decay $\bar{B}^0 \to D^\ast\tau^-\bar\nu_{\tau}$. The thick black dashed lines are the SM prediction; the gray bands include NP effects corresponding to the $2\sigma$ allowed regions in Fig.~\ref{fig:constraint}; the red dotted lines represent the best-fit values of the NP couplings given in Eq.~(\ref{eq:bestfit}).}
\label{fig:pol-BDv}
\end{figure}

The $\tau^-$ polarization components in $\bar{B}^0 \to D^\ast\tau^-\bar\nu_{\tau}$ are shown in Fig.~\ref{fig:pol-BDv}. In each column we present one component in the presence of different NP couplings $S_L$, $V_R$, and $T_L$, one by one. In each row one can see how one NP coupling affects the three components at the same time. All the plots are in one scale so that one can quickly compare the sensitivity of different polarization components to different NP couplings. 

Let us begin with the longitudinal polarization (left column in Fig.~\ref{fig:pol-BDv}). The longitudinal polarization $P_L^{D^\ast}$ is not affected by $V_R$ but is very sensitive to $S_L$ and $T_L$. Both $S_L$ and $T_L$ tend to increase $P_L^{D^\ast}$ and shift the zero-crossing point from that in the SM. In the presence of $S_L$, $P_L^{D^\ast}$ starts at a higher value but converges to its SM value at high $q^2$ and its shape is similar to the SM one. In contrast to $S_L$, $T_L$ changes $P_L^{D^\ast}$ thoroughly: $P_L^{D^\ast}$ now starts at a lower position but can be positive for the most part of the whole $q^2$ region and maximally diverts from its SM prediction at high $q^2$.

The transverse polarization $P_T^{D^\ast}$ (center column in Fig.~\ref{fig:pol-BDv}) has the same sensitivity to $S_L$ and $V_R$ but $S_L$ tends to increase $P_T^{D^\ast}$ while $V_R$ tends to decrease $P_T^{D^\ast}$. The transverse polarization is extremely sensitive to $T_L$ and its sign can be changed in the presence of $T_L$. It is interesting to note that $S_L$
increases both $P_L^{D^\ast}$ and $P_T^{D^\ast}$, while $T_L$ amplifies $P_L^{D^\ast}$ but lowers $P_T^{D^\ast}$. When $T_L$ is present, largest deviations of $P_T^{D^\ast}$ from its SM prediction happen at low $q^2$, which is opposite to the case of $P_L^{D^\ast}$.

Regarding the normal polarization $P_N^{D^\ast}$ (right column in Fig.~\ref{fig:pol-BDv}), it is sensitive to both $S_L$ and $T_L$ but slightly more to $S_L$. $P_N^{D^\ast}$ can be both positive or negative and its absolute value can reach about $0.2$. It is worth noting that $P_N^{D^\ast}$ is much less sensitive to $T_L$ in comparison with $P_L^{D^\ast}$ and $P_T^{D^\ast}$.

\begin{figure}[htbp]
\begin{tabular}{lll}
\includegraphics[scale=0.4]{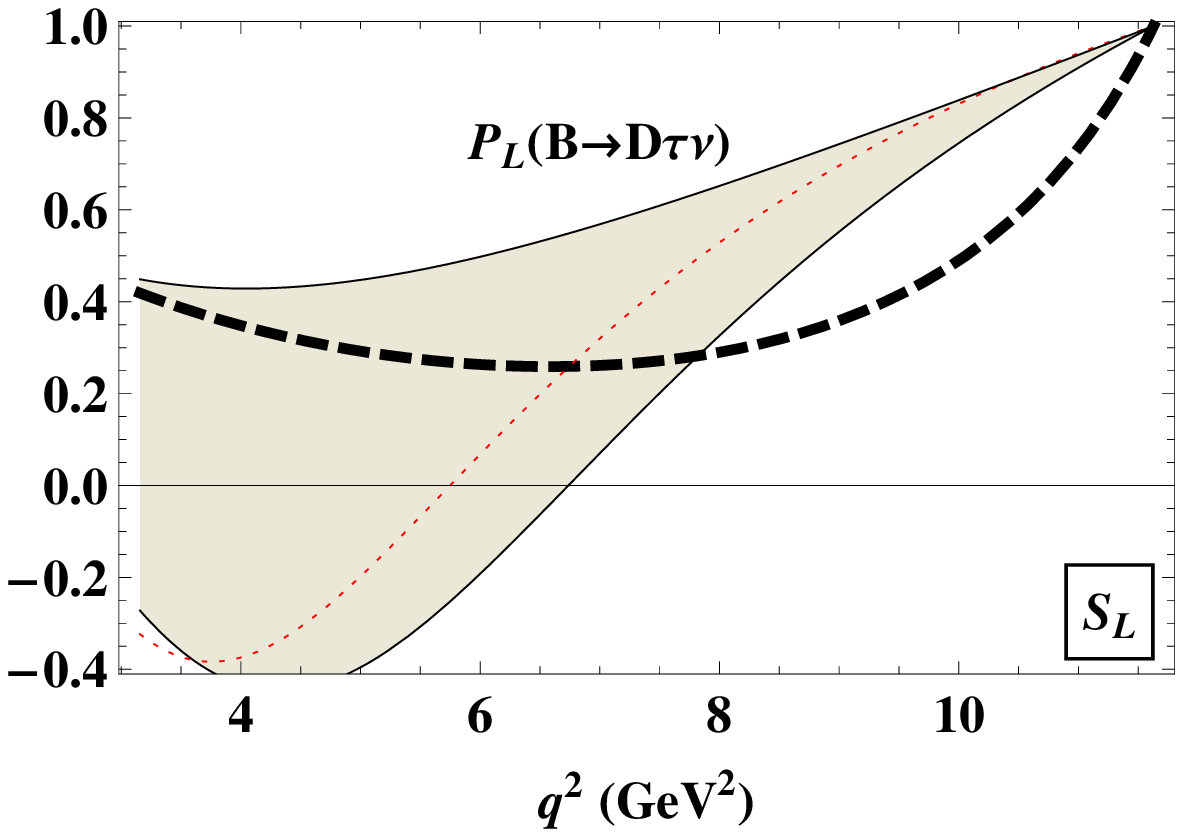}
& 
\includegraphics[scale=0.4]{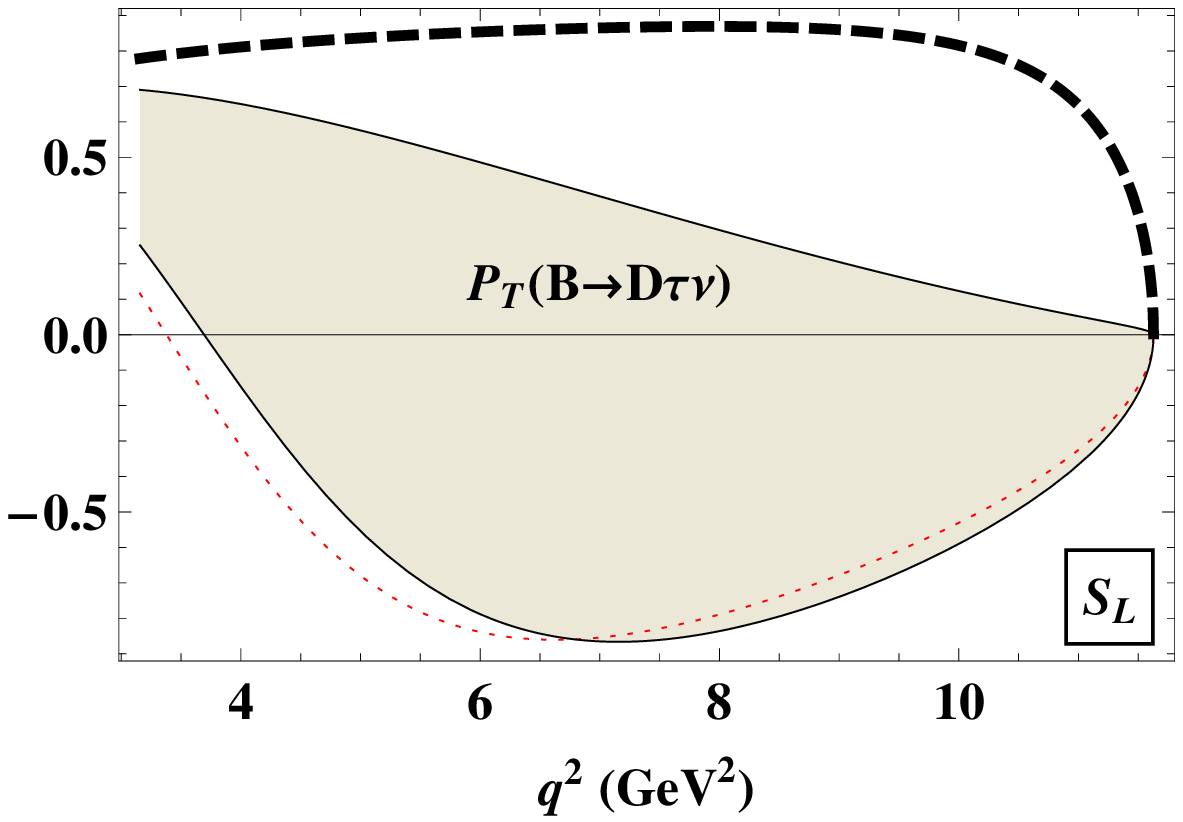}
& 
\includegraphics[scale=0.4]{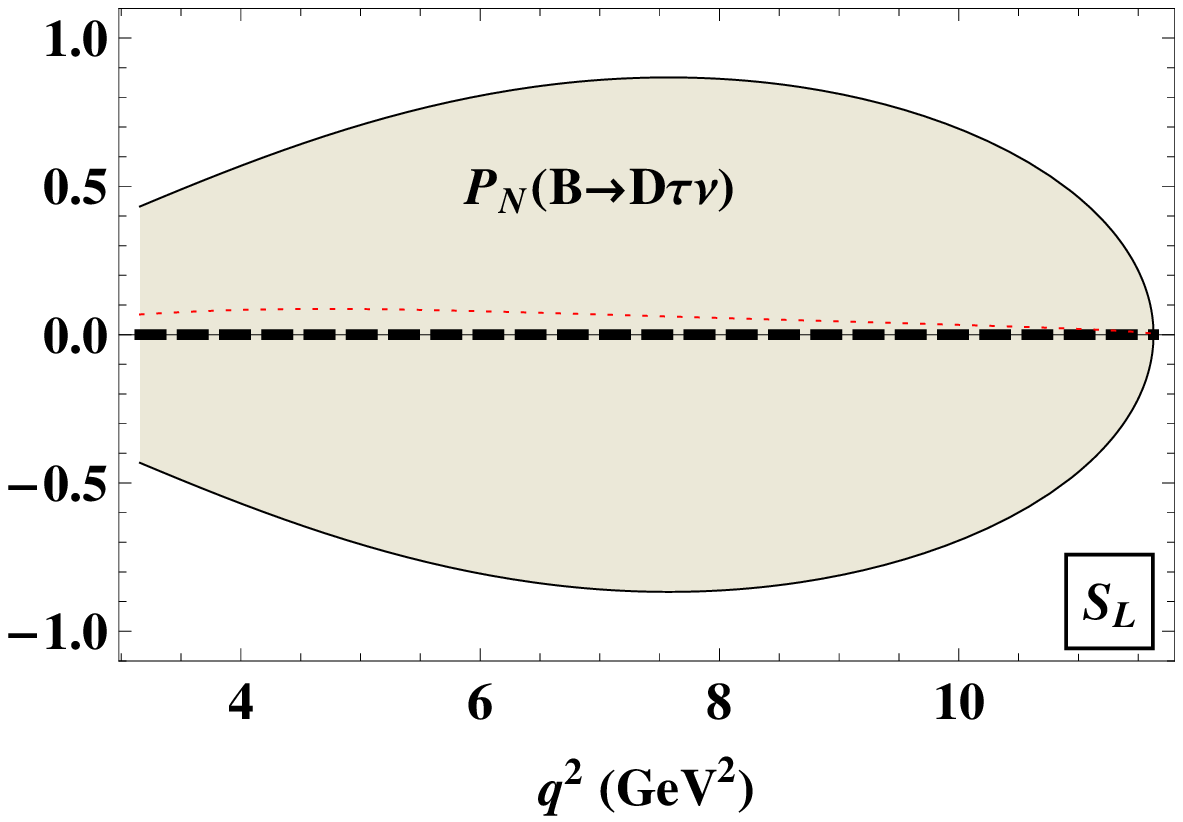}\\
\includegraphics[scale=0.4]{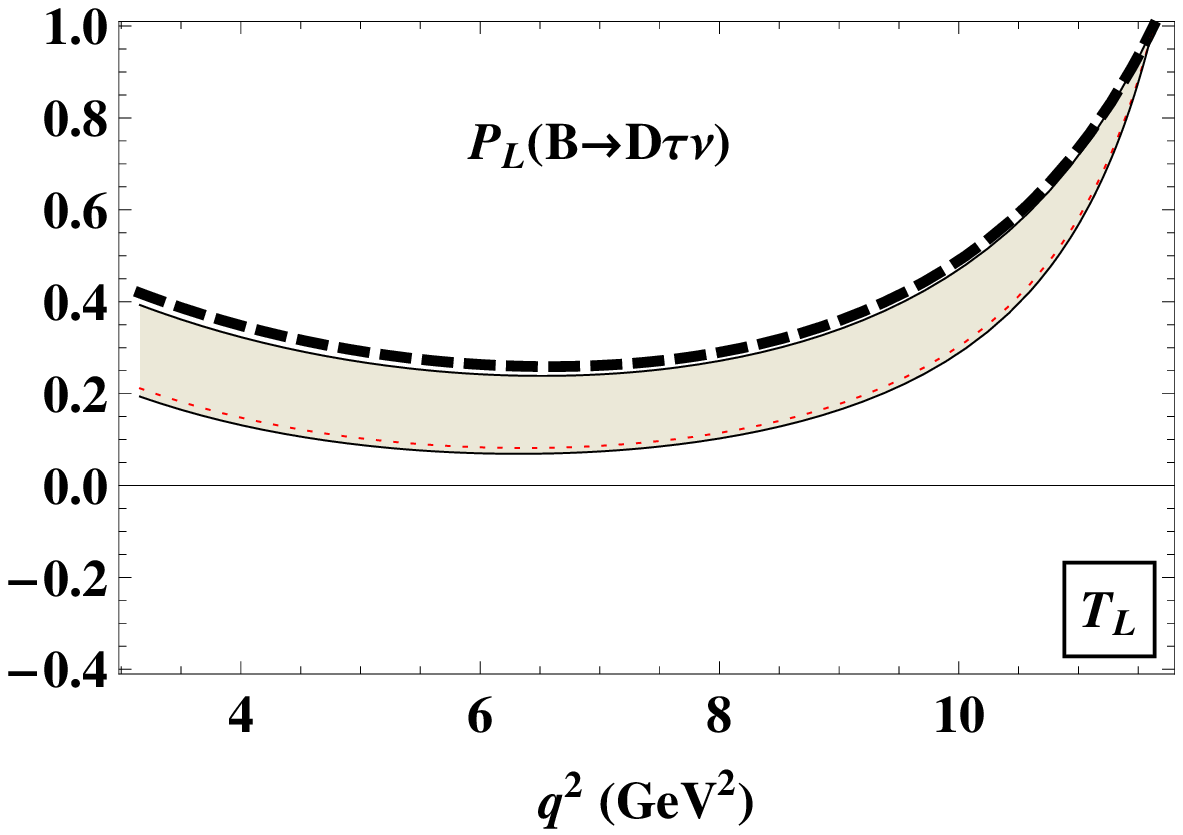}
& 
\includegraphics[scale=0.4]{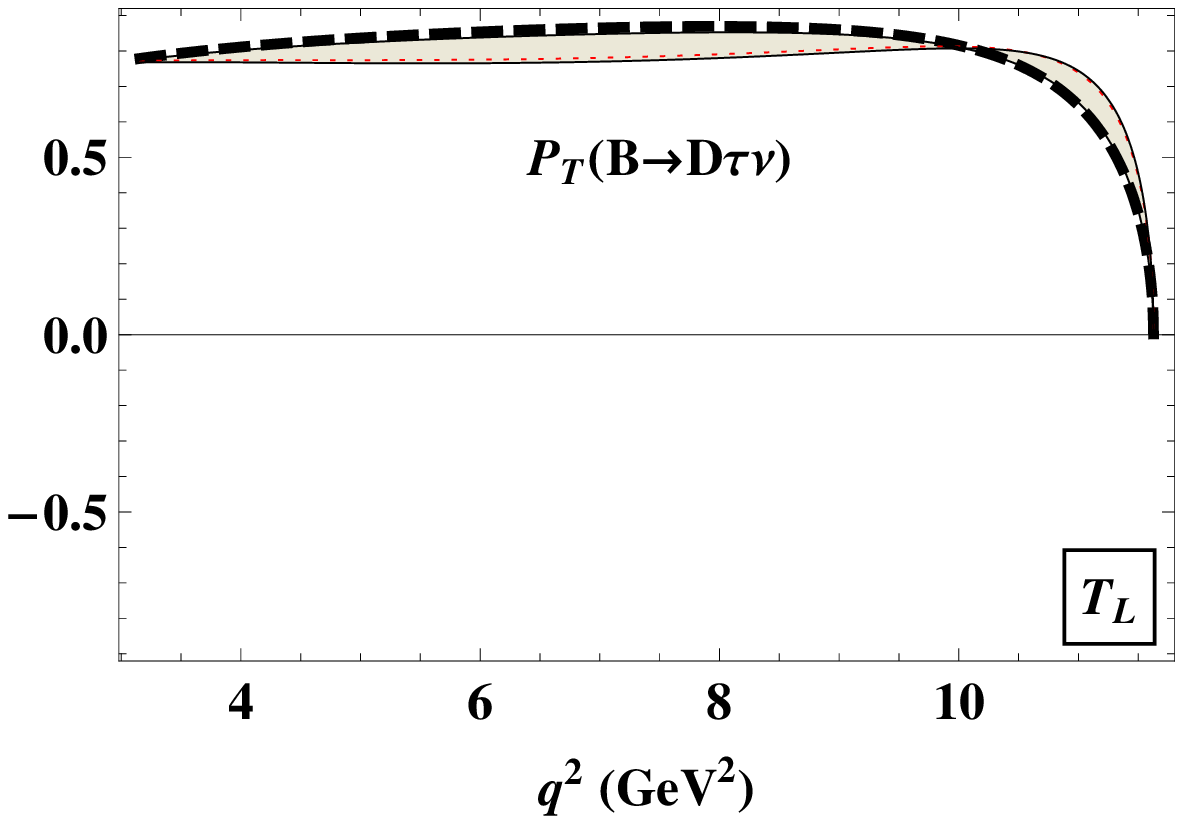}
& 
\includegraphics[scale=0.4]{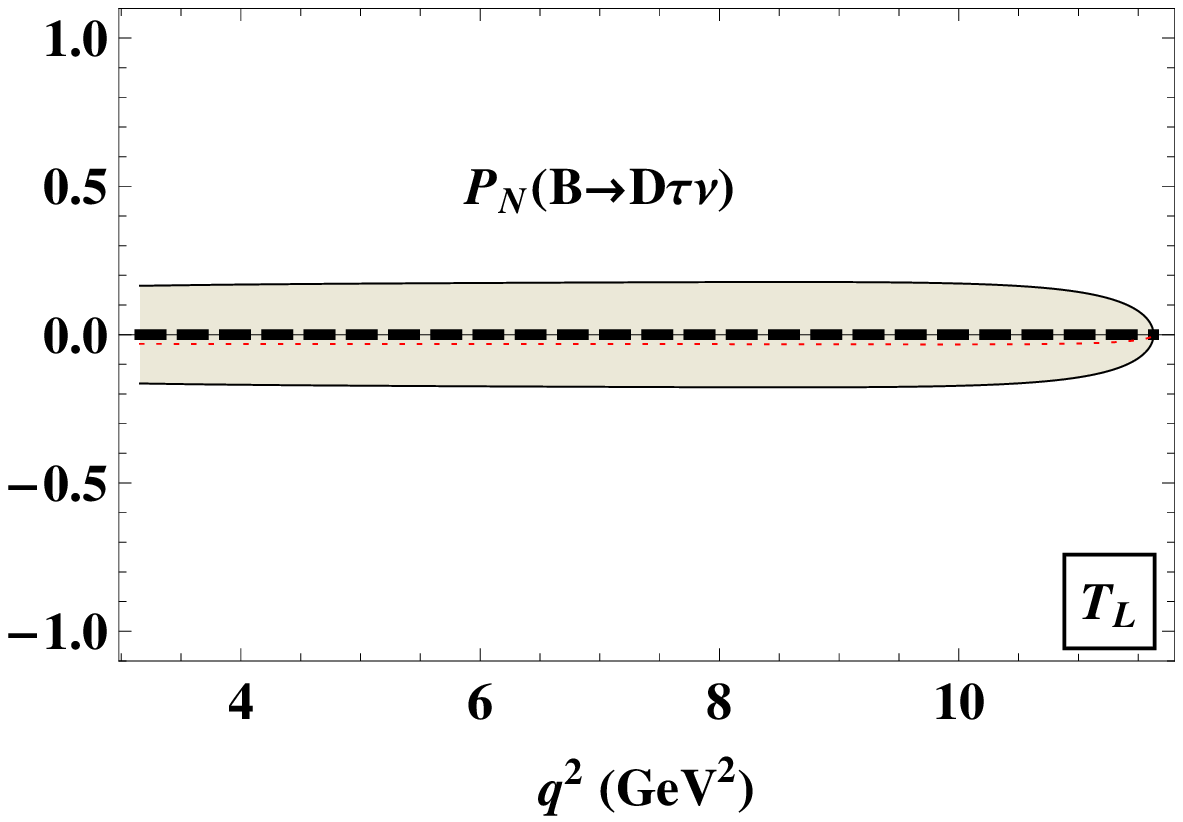}
\end{tabular}
\caption{Longitudinal (left), transverse (center), and normal (right) polarization of the $\tau^-$ in the decay $\bar{B}^0 \to D\tau^-\bar\nu_{\tau}$. Notations are the same as in Fig.~\ref{fig:pol-BDv}.}
\label{fig:pol-BD}
\end{figure}

Next we turn to the $\tau^-$ polarizations in $\bar{B}^0 \to D\tau^-\bar\nu_{\tau}$, which are shown in Fig.~\ref{fig:pol-BD}. 
It is readily seen that all three polarization components in this case are much more sensitive to $S_L$ than to $T_L$. In the presence of $T_L$, the polarizations $P_L^D$ and $P_T^D$ can be positively or negatively enhanced but their shapes over the whole $q^2$ range are similar to those in the SM. In contrast, the scalar coupling $S_L$ changes the shapes of $P_L^D$ and $P_T^D$ dramatically and can even imply a zero-crossing point, which is impossible in the SM. This distinct effect of $S_L$ may give some hints for experimental study. The normal polarization $P_N^D$ can reach about $\pm0.2$ under the effect of $T_L$ while it can even reach about $\pm0.8$ when $S_L$ is present. 

The $q^2$ dependence of the polarizations bears powerful information for discriminating between different NP scenarios. One possible approach is to make use of it to perform a bin-by-bin analysis in order to probe NP in different $q^2$ regions. One can also calculate the average polarizations over the whole $q^2$ region. When calculating the $q^{2}$ averages one has to multiply the numerator and denominator of~(\ref{eq:PL}),~(\ref{eq:PT}), and~(\ref{eq:PN}) by the 
$q^{2}$-dependent piece of the phase-space factor given by  
$
C(q^2) = |\mathbf{p_2}| (q^2-m_\tau^2)^2/q^2,
$
where  $|{\bf p_2}|=\lambda^{1/2}(m_1^2,m_2^2,q^2)/2m_1 $ is the momentum of the daughter meson.
For example, the average longitudinal polarization $\langle P_L^D\rangle$ can then be calculated  according to
\be
\langle P_L^D\rangle = 
\frac{\int dq^{2} C(q^{2})\big(P_L^D(q^2){\cal H}_{\rm tot}^D\big)}
{\int dq^{2} C(q^{2}){\cal H}_{\rm tot}^D}.
\label{eq:PLDint}
\en
The predictions for the mean polarizations are summarized in Table~\ref{tab:pol-average}. 
\begin{table}[htbp] 
\begin{center}
\begin{tabular}{lccccc}
\hline\hline
\multicolumn{4}{c}{ $\bar{B}^0\to D$ }\\
\hline
&\quad  $<P_{L}^D>$\qquad 
&\quad  $<P_{T}^D>$ \qquad   
&\quad  $<P_{N}^D>$ \qquad  
&\quad  $<|\vec P^D|>$ \qquad
\\
\hline
SM (CCQM) &\quad $0.33$\quad &\quad $0.84$ \quad &\quad $0$\quad&\quad $0.91$\quad\\
$S_L$ 
&\quad $(0.36,0.67)$\quad
&\quad $(-0.68,0.33)$\quad
&\quad $(-0.76,0.76)$\quad
&\quad $(0.89,0.96)$\quad
\\ 
$T_L$
&\quad $(0.13,0.31)$\quad
&\quad $(0.78,0.83)$\quad
&\quad $(-0.17,0.17)$\quad
&\quad $(0.79,0.90)$\quad\\
\hline\hline
\multicolumn{4}{c}{ $\bar{B}^0\to D^\ast$ }\\
\hline
&\quad  $<P_{L}^{D^\ast}>$ \qquad 
&\quad  $<P_{T}^{D^\ast}>$ \qquad   
&\quad  $<P_{N}^{D^\ast}>$ \qquad  
&\quad $<|\vec P^{D^\ast}|>$\qquad
\\
\hline
SM (CCQM)&\quad -0.50\quad &\quad 0.46\quad &\quad 0 \quad &\quad 0.71\quad\\
$S_L$
&\quad $(-0.40,-0.14)$\quad
&\quad $(0.47,0.62)$\quad
&\quad $(-0.20,0.20)$\quad
&\quad $(0.69,0.70)$\quad
\\
$T_L$
&\quad $(-0.36,0.24)$\quad
&\quad $(-0.61,0.26)$\quad
&\quad $(-0.17,0.17)$\quad
&\quad $(0.23,0.69)$\quad
\\
$V_R$
&\quad $-0.50$\quad
&\quad $(0.32,0.43)$\quad
&\quad $0$\quad
&\quad $(0.48,0.67)$\quad
\\
\hline\hline
\end{tabular}
\caption{$q^{2}$ averages of the polarization components and the total polarization. The two rows labeled by SM (CCQM) contain our predictions using SM effective operators with transition form factors calculated in the CCQM. The predicted intervals for the observables in the presence of NP are given in correspondence with the $2\sigma$ allowed regions of the NP couplings depicted in Fig.~\ref{fig:constraint}.}
\label{tab:pol-average}
\end{center}
\end{table}
Again, one sees that the $\tau^-$ polarization components in $\bar{B}^0 \to D\tau^-\bar\nu_{\tau}$ are extremely sensitive to $S_L$. When $S_L$ is present, $\langle P_L^D\rangle$ can be as large as $0.67$, $\langle P_T^D\rangle$ can reach $-0.68$, and $\langle P_N^D\rangle$ can even reach $\pm 0.76$. It is interesting to note that if one measures $\langle P_L^D\rangle$ and finds any excess over the SM value, it would be a clear sign of $S_L$. Meanwhile, the $\tau^-$ longitudinal and transverse polarization components in $\bar{B}^0 \to D^\ast\tau^-\bar\nu_{\tau}$ are more sensitive to $T_L$. The coupling $T_L$ can enhance $\langle P_L^{D^\ast}\rangle$ from the SM value of $-0.50$ up to $0.24$, or lower $\langle P_T^{D^\ast}\rangle$ from $0.46$ down to $-0.61$. Notably, the average transverse polarization $\langle P_T^D\rangle$ is almost insensitive to $T_L$ in comparison with $S_L$. When $T_L$ is present, one finds $0.78\leq\langle P_T^D\rangle\leq 0.83$, which is almost the same as the SM value $\langle P_T^D\rangle=0.84$. In contrast, if $S_L$ is present, one has $-0.68\leq\langle P_T^D\rangle\leq 0.33$, which is much lower than the SM prediction. This unique property of $\langle P_T^D\rangle$ may play a very important role in probing the scalar coupling $S_L$. It is also interesting to note that the average total polarization $<|\vec P^{D^\ast}|>$ is almost insensitive to $S_L$. 
\section{Summary and conclusions}
\label{sec:summary}
We have studied the longitudinal, transverse, and normal polarization components of the $\tau^-$ in the semileptonic decays $\bar{B}^0 \to D^{(\ast)}\tau^-\bar\nu_{\tau}$ in the presence of NP scalar, vector, and tensor interactions based on an SM-extended effective Hamiltonian. Constraints on the space of NP couplings have been obtained from experiments at $B$ factories and LHCb including the most recent result of the Belle collaboration~\cite{Hirose:2016wfn}. We have also briefly discussed how to extract the polarization of the $\tau^-$ from the distribution of its most prominent
subsequent decay modes. 

All the polarization components are sensitive to the scalar coupling $S_L$ and the tensor coupling $T_L$. Besides, the transverse polarization $P_T^{D^\ast}$ is also sensitive to the vector coupling $V_R$. The longitudinal and transverse polarizations are more sensitive to $T_L$ in the case of $\bar{B}^0 \to D^\ast\tau^-\bar\nu_{\tau}$, but more to $S_L$ in the case of $\bar{B}^0 \to D\tau^-\bar\nu_{\tau}$. $P_N^{D^\ast}$ is equally sensitive to $T_L$ and $S_L$, while $P_N^D$ is much more sensitive to $S_L$ than to $T_L$. The normal polarization can reach about $\pm0.8$ in $\bar{B}^0 \to D\tau^-\bar\nu_{\tau}$ if $S_L$ is present, and about $\pm0.2$ in other cases. These observations may provide some insights to look for NP in the decays $\bar{B}^0 \to D^{(\ast)}\tau^-\bar\nu_{\tau}$.

\begin{acknowledgments}
  The authors thank the Heisenberg-Landau Grant for providing support for their
  collaboration. M.A.I. acknowledges the financial support of the PRISMA Cluster of
  Excellence at the University of Mainz.  M.A.I. and C.T.T. greatly appreciate the warm
  hospitality of the Mainz Institute for Theoretical Physics (MITP) at the University of Mainz.
\end{acknowledgments}

\appendix*
\section{Helicity amplitudes}
\label{app:helicity}
In this appendix, we express the helicity amplitudes used in the main text
in terms of the invariant form factors defined in Eq.~(\ref{ff}). A detailed
derivation of these relations can be found in our recent
paper~\cite{Ivanov:2016qtw}. 

For the $\bar{B}^0\to D$ transition:
\bea
H_t   = \frac{Pq F_+ + q^2 F_-}{\sqrt{q^2}},\, 
H_0 =\frac{2m_1|{\bf p_2}|F_+}{\sqrt{q^2}},\,
H_P^S = (m_1+m_2)F^S,\,
H_T =\frac{2m_1|{\bf p_2}|F^T}{m_1+m_2},
\label{eq:hel_pp}
\ena
where $|{\bf p_2}|=\lambda^{1/2}(m_1^2,m_2^2,q^2)/2m_1$ is the momentum of the daughter meson. 

For the $\bar{B}^0\to D^\ast$ transition:
\bea
H_{0} &=&\frac{-Pq(m_1^2 - m_2^2 - q^2)A_0 + 4m_1^2|{\bf p_2}|^2 A_+}{2m_2\sqrt{q^2}(m_1+m_2)} 
,\nn
H_{t} &=&\frac{m_1|{\bf p_2}|\left(Pq(-A_0+A_+)+q^2 A_-\right)}{m_2\sqrt{q^2}(m_1+m_2)},
\nn
H_{\pm} &=& \frac{-Pq A_0\pm 2m_1|{\bf p_2}| V}{m_1+m_2},\nn
H^S_V &=& \frac{m_1}{m_2}|{\bf p_2}|G^S, \nn
H_T^\pm &=& -\frac{1}{\sqrt{q^2}}\left[
(m_1^2-m_2^2\pm 2m_1|{\bf p_2}|)G_1^T+q^2G_2^T
\right],\nn
H_T^0 &=& -\frac{1}{2m_2}
\Big[
(m_1^2+3m_2^2-q^2)G_1^T +(m_1^2-m_2^2-q^2)G_2^T-\frac{4m_1^2|{\bf p_2}|^2}{(m_1+m_2)^2}G_0^T
\Big].
\ena
The dependence of the helicity amplitudes and the invariant form factors on $q^2$ have been omitted for simplicity.

 \ed